\documentclass[12pt]{article}

\begin{document}
\hoffset=-0.75in

\title{$PT$ symmetry as a necessary and sufficient condition for unitary time evolution}

\author{Philip~D.~Mannheim \\ Department of Physics, University of Connecticut, Storrs, CT
06269, USA
\\ email: philip.mannheim@uconn.edu}

\date{November 5, 2012}

\maketitle

While Hermiticity of a time-independent Hamiltonian leads to unitary time evolution, in and of itself, the requirement of Hermiticity is only sufficient for unitary time evolution. In this paper we provide conditions that are both necessary and sufficient. We show that $PT$ symmetry of a time-independent Hamiltonian, or equivalently, reality of the secular equation that determines its eigenvalues, is both necessary and sufficient for unitary time evolution. For any $PT$-symmetric Hamiltonian $H$ there always exists an operator $V$ that relates $H$ to its Hermitian adjoint according to $VHV^{-1}=H^{\dagger}$. When the energy spectrum of $H$ is complete, Hilbert space norms $\langle\psi_1|V|\psi_2\rangle$ constructed with this $V$ are always preserved in time. With the energy eigenvalues of a real secular equation being either real or appearing in complex conjugate pairs, we thus establish the unitarity of time evolution in both cases. We also establish the unitarity of time evolution for Hamiltonians whose energy spectra are not complete. We show that when the energy eigenvalues of a Hamiltonian are real and complete the operator $V$ is a positive Hermitian operator, which has an associated square root operator that can be used to bring the Hamiltonian to a Hermitian form. We show that systems with $PT$-symmetric Hamiltonians obey causality. We note that Hermitian theories are ordinarily associated with a path integral quantization prescription in which the path integral measure is real, while in contrast non-Hermitian but $PT$-symmetric theories are ordinarily associated with path integrals in which the measure needs to be complex, but in which the Euclidean time continuation of the path integral is nonetheless real. Just as  the second-order Klein-Gordon theory is stabilized against transitions to negative frequencies because its Hamiltonian is positive definite, through $PT$ symmetry the fourth-order derivative Pais-Uhlenbeck theory can equally be stabilized.

\medskip


\section{Unitary time evolution for non-Hermitian Hamiltonians}
\label{S1}

If for a time-independent Hamiltonian $H$ one constructs a time evolution operator of the form $U(t)=e^{-iHt}$, then if $H$ is Hermitian, the adjoint operator $U^{\dagger}(t)$ is given by $U^{\dagger}(t)=e^{iHt}$ and the unitarity of $U(t)$ immediately follows. Moreover, if one introduces right-eigenvectors of $H$ that obey 
\begin{equation}
i\frac{d}{dt} |R_i(t)\rangle=H|R_i(t)\rangle,
\label{E1}
\end{equation}
then for Hermitian Hamiltonians the Dirac norm $\langle R_i(t)|R_i(t)\rangle$ is immediately preserved in time. However, if the Hamiltonian is not Hermitian, conjugation of (\ref{E1}) leads to 
\begin{equation}
-i\frac{d}{dt} \langle R_i(t)|=\langle R_i(t)|H^{\dagger},
\label{E2}
\end{equation}
with $U^{\dagger}(t)$ now being given by $U^{\dagger}(t)=e^{iH^{\dagger}t}$, and with the Dirac norm then not being time independent. Despite this, one cannot conclude that the time evolution associated with a non-Hermitian Hamiltonian is not unitary. Rather, all one can conclude is that the standard derivation of unitarity for Hermitian Hamiltonians and the appropriateness of the Dirac norm that goes hand in hand with it do not apply when the Hamiltonian is not Hermitian.  While the above analysis shows that Hermiticity implies unitarity, one cannot conclude that lack of Hermiticity implies lack of unitarity. And indeed, a whole class of non-Hermitian Hamiltonians has been identified for which time evolution is unitary in the sense that for them one can find appropriate norms that are preserved in time (this being the sense in which the term ``unitary time evolution" is used in the $PT$ literature), with the class being those particular non-Hermitian Hamiltonians that possesses a $PT$ symmetry and have energy eigenvalues that are all real (see the review of  \cite{R1} and references therein). The purpose of this paper, which in part serves as a review of work in the field, is to identify all possible classes of Hamiltonians for which time evolution is unitary in this sense, with Hermitian Hamiltonians being just one of the possible classes.

The key to finding the general rule is to recognize that if $|R_i(t)\rangle$ is an eigenstate of $H$ with some general complex energy $E_i=E^R_i+iE^I_i$, then on introducing some general time-independent operator $V$, the $V$-dependent inner product $\langle R_j(t)|V|R_i(t)\rangle$ will be given by
\begin{equation}
\langle R_j(t)|V|R_i(t)\rangle = e^{-i(E^R_i-E^R_j)t+(E^I_i+E^I_j)t}\langle R_j(t=0)|V|R_i(t=0)\rangle.
\label{E3}
\end{equation}
This inner product will be time independent for energies that obey the two conditions 
\begin{equation}
E^R_i-E^R_j=0, \qquad E^I_i+E^I_j=0, 
\label{E4}
\end{equation}
and will be time independent for states with energies that do not obey these conditions if an operator $V$ can be found so that for all of those states the matrix element $\langle R_j(t=0)|V|R_i(t=0)\rangle$ vanishes. The eigenstates of Hermitian Hamiltonians immediately obey these conditions with $E^I_i=0$, $E^I_j=0$ and $V=I$. However in the general case one needs a non-trivial $V$, and a thus non-Dirac norm.

To find the general conditions under which the needed $V$ will in fact exist, we note that from (\ref{E1}) and (\ref{E2}) we can evaluate the time derivative of $\langle R_j(t)|V|R_i(t)\rangle$ to obtain
\begin{equation}
i\frac{d}{dt} \langle R_j(t)|V|R_i(t)\rangle=\langle R_j(t)|(VH-H^{\dagger}V)|R_i(t)\rangle.
\label{E5}
\end{equation}
If now we require that the set of states $|R_i(t)\rangle$ be complete for the general $H$, Hermitian or otherwise, then the time independence of all  $\langle R_j(t)|V|R_i(t)\rangle$ matrix elements will be secured if $V$ obeys the relation
\begin{equation}
VHV^{-1}=H^{\dagger}
\label{E6}
\end{equation}
as an operator identity.\footnote{We discuss below what happens when eigenstates are not complete.}  The needed $V$ is thus the one that effects the transformation from $H$ to its Hermitian adjoint $H^{\dagger}$. Conversely, if a Hamiltonian has the property that $H$ and $H^{\dagger}$ can be related by an operator $V$ as in (\ref{E6}), it follows from (\ref{E5}) that all matrix elements $\langle R_j(t)|V|R_i(t)\rangle$ constructed with this $V$ are time independent. The condition that there exists an operator $V$ that effects $VHV^{-1}=H^{\dagger}$ is thus a necessary and sufficient condition for unitary time evolution. We note that as defined in (\ref{E6}) the operator $V$ depends on the Hamiltonian. Thus unlike the standard Dirac norm $\langle R_j(t)|R_i(t)\rangle$, the $\langle R_j(t)|V|R_i(t)\rangle$ norm cannot be preassigned, with it being dynamically determined by the Hamiltonian itself.\footnote{In this respect the situation is analogous to the difference between special relativity and general relativity. In special relativity the spacetime Minkowski metric is preassigned, while in general relativity the spacetime metric is dynamically determined. Moreover, the analogy even extends to the difference between geometric metrics with Euclidean and Minkowski signature. Specifically, flat space geometric metrics with positive signature are the analog of the standard Dirac Hilbert space norm, while flat space metrics with Minkowski signature are the analog of an indefinite Dirac norm. Curved space geometric metrics with positive signature are the analog of the $V$-type Hilbert space norm discussed here, while curved space metrics with Minkowski signature are the analog of an indefinite $V$-type norm.} In addition, we note that for any Hamiltonian that does obey (\ref{E6}), both it and its Hermitian adjoint must have the same set of eigenvalues. Such eigenvalues can thus only be real or appear in complex conjugate pairs. In our discussion below of $PT$-symmetric Hamiltonians we will encounter precisely this same requirement on energy eigenvalues. 

When there does exist an operator  $V$ for which (\ref{E6}) is satisfied, we can rewrite (\ref{E2}) as 
\begin{equation}
-i\frac{d}{dt} \langle R_i(t)|V=\langle R_i(t)|VH.
\label{E7}
\end{equation}
In such a situation the state $\langle L_i(t)|=\langle R_i(t)|V$ is a left-eigenvector of $H$ as it obeys
\begin{equation}
-i\frac{d}{dt} \langle L_i(t)|=\langle L_i(t)|H.
\label{E8}
\end{equation}
In terms of the left-eigenvectors the $\langle R_j(t)|V|R_i(t)\rangle$ inner product can be written as $\langle L_j(t)|R_i(t)\rangle$, with the inner product thus being the overlap of the left- and right-eigenvectors of the Hamiltonian. Because of (\ref{E1}) and (\ref{E8}) this inner product obeys $\langle L_j(t)|R_i(t)\rangle=\langle L_j(0)|e^{-iHt}e^{iHt}|R_i(0)\rangle=\langle L_j(0)|R_i(0)\rangle$ and is manifestly time independent. 

Recalling that the standard Dirac norm  $\langle R_j(t)|R_i(t)\rangle=\langle R_j(0)|e^{-iH^{\dagger}t}e^{iHt}|R_i(0)\rangle$ is time independent (and equal to $\langle R_j(0)|R_i(0)\rangle$) when the Hamiltonian is Hermitian, and recalling that unitarity is secured through the completeness of energy eigenstates, we see that though use of  $\langle L_j(t)|R_i(t)\rangle=\langle R_j(t)|V|R_i(t)\rangle$ norm and a completeness requirement on energy eigenstates we can generalize the concept of unitarity to the non-Hermitian case.  Thus even in the non-Hermitian case we can secure time-independent evolution of inner products. Now regardless of whether or not a Hamiltonian might be Hermitian, the Hamiltonian is always  the generator of time translations and $U(t)=e^{-iHt}$ is always the time evolution operator. For a non-Hermitian Hamiltonian that obeys  (\ref{E6}) the time evolution operator thus obeys
\begin{equation}
U^{\dagger}(t)=e^{iH^{\dagger}t}=Ve^{iHt}V^{-1}=VU^{-1}(t)V^{-1}.
\label{E8p}
\end{equation}
Thus for non-Hermitian Hamiltonians that obey (\ref{E6}) unitarity is understood to mean that $U^{\dagger}(t)$ is not equal to $U^{-1}(t)$ but to its $V$ transform as given in (\ref{E8p}). Below in Sec. (3) we shall show how to formulate unitarity in the scattering matrix language when the Hamiltonian is not Hermitian. 

Now while we have defined a generalized inner product as  $\langle R_j(t)|V|R_i(t)\rangle$ in the non-Hermitian case, to be able to associate it with a conventional positive-definite quantum-mechanical probability amplitude we would additionally need to require that all energy eigenvalues be real and complete. Specifically, as we show in Sec. (3), in the case where all energy eigenvalues are real and complete,  we can bring the Hamiltonian $H$ to a Hermitian form $\tilde{H}$  via a similarity transform. In such a situation the $\langle R_j(t)|V|R_i(t)\rangle$ norm is transformed into the standard positive-definite Dirac norm $\langle \tilde{R}_j(t)|\tilde{R}_i(t)\rangle=\delta_{ij}$ associated with the Hermitian $\tilde{H}$, and thus has the same positive-definite probability amplitude interpretation. Inner products associated with non-Hermitian Hamiltonians that are similarity equivalent to Hermitian Hamiltonians thus have a conventional quantum-mechanical probability interpretation.

To explicitly construct a coordinate space representation for the probability amplitude we need to represent the Hamiltonian as a differential operator in coordinate space. In the illustrative one-dimensional case we introduce left and right wave functions that obey $i\partial_t\psi^{{\rm R}}_i(x, t)=\stackrel{\rightarrow}{H}\psi^{{\rm R}}_i(x,t)$ with $H$ acting to the right, and $-i\partial_t\psi^{{\rm L}}_i(x,t)=\psi^{{\rm L}}_i(x,t)\stackrel{\leftarrow}{H}$ with $H$ acting to the left.\footnote{To construct operators that act to the left in wave mechanics one uses a realization of the $[x,p]=i$ commutator in which the momentum acts  as $+i\partial_x$.} For these left and right wave functions we obtain   
\begin{equation}
i\partial_t\left[\int dx \psi^{{\rm L}}_i(x, t)\psi^{{\rm R}}_j(x, t)\right]=\int dx \left[\psi^{{\rm L}}_i(x, t)\stackrel{\rightarrow}{H}\psi^{{\rm R}}_j(x, t)-\psi^{{\rm L}}_i(x, t)\stackrel{\leftarrow}{H}\psi^{{\rm R}}_j(x, t)\right].
\label{E8q}
\end{equation}
Thus when the integrand on the right-hand side of (\ref{E8q}) is a total spatial derivative, asymptotic boundedness of the wave functions secures both the time independence of $\int dx \psi^{{\rm L}}_i(x, t)\psi^{{\rm R}}_j(x, t)$ and its finiteness. If all energy eigenvalues are real, the overlap integral $\int dx \psi^{{\rm L}}_i(x, t)\psi^{{\rm R}}_j(x, t)$ will be proportional to  $\delta_{ij}$ with a phase $\eta_j$. However, with the overlap integral not being that of  a wave function with its own Hermitian conjugate, the overlap integral is not constrained to be positive definite, and $\eta_j$ could thus be positive or negative, or even complex. If all energy eigenvalues are real then, the overlap integral can be written as 
\begin{equation}
\int dx \psi^{{\rm L}}_i(x, t)\psi^{{\rm R}}_j(x, t)=\eta_i\delta_{i,j}
\label{E8r}
\end{equation}
with $\eta_j$ as yet undetermined.

In terms of left and right energy eigenstates of $H$ it would be very convenient if we could fix the phases of  $\psi^{{\rm R}}_i(x, t)$ and $\psi^{{\rm L}}_i(x, t)$ so that we could identify $\psi^{{\rm R}}_i(x, t)=\langle x|R_{i}(t)\rangle$,  $\psi^{{\rm L}}_i(x, t)=\eta_i\langle L_{i}(t)|x\rangle$. With such a choice of phases we would then obtain 
\begin{equation}
\int dx \psi_i^{{\rm L}}(x, t)\psi_j^{{\rm R}}(x, t)=\int dx \langle L_{i}(t)|x\rangle \eta_i\langle x|R_{j}(t)\rangle
= \eta_i\langle L_{i}(t)|R_{j}(t)\rangle.
\label{E8s}
\end{equation}
Now we recall that in the event that eigenspectrum of $H$ is real and complete we can set $\langle L_j(t)|R_i(t)\rangle=\langle R_j(t)|V|R_i(t)\rangle=\delta_{ij}$, and can thus recover (\ref{E8r}). The utility of doing this is that when the eigenspectrum of a $PT$-symmetric Hamiltonian is complete one is able to introduce a $C$ operator that obeys  $[C,H]=0$, $C^2=I$ \cite{R1}. Such a $C$ operator can be diagonalized simultaneously with $H$, with its eigenvalues being given by $C_i=\pm 1$. With such a $C$ operator we can set the phase of $\psi^{{\rm L}}_i(x, t)$ so that $\psi^{{\rm L}}_i(x, t)=\langle L_{i}(t)|C|x\rangle=C_i\langle L_{i}(t)|x\rangle$, and can thus identify $\eta_i$ with $C_i$, with $\eta_i$ now being restricted to the values plus or minus one. In general then, when there is a similarity equivalence to a Hermitian theory the appropriate $PT$ theory probability amplitude is given by  the positive definite $\int dx \psi_i^{{\rm L}}(x, t)C_i\psi_i^{{\rm R}}(x, t)$.

In the event that some of the energy eigenvalues of a non-Hermitian Hamiltonian are not real but instead come in complex conjugate pairs, the relevant $\langle R_j(t)|V|R_i(t)\rangle$ matrix elements would not then need to be positive definite as they would then be associated with transitions between eigenstates of complex conjugate energy eigenvalues. Such transitions are discussed further in Sec. (4). To summarize then, to obtain unitary time evolution the general rule is to define the inner product as the overlap of the left- and right-eigenvectors, and so we need to ascertain the conditions on $H$ for which a $V$ operator that satisfies (\ref{E6}) (and thus the left-eigenvectors that it then generates) will necessarily  exist.

\section{$PT$ symmetry and the $VHV^{-1}=H^{\dagger}$ condition}
\label{S2}

When $H$ and $H^{\dagger}$ are related according to the similarity transform $VHV^{-1}=H^{\dagger}$, the secular equation for the eigenvalues of $H$ is of the form ${\rm det}(H-\lambda I)={\rm det}(V^{-1}H^{\dagger}V-\lambda I)={\rm det}(H^{\dagger}-\lambda I)={\rm det}((H^{*})^{\rm TR}-\lambda I)={\rm det}(H^{*}-\lambda I)=0$, where ${\rm TR}$ denotes transpose. Consequently, $H$ and $H^{*}$ possess the same set of eigenvalues, with the secular equation ${\rm det}(H-\lambda I)=0$ thus being real. With the energies of a real secular equation either being real or appearing in complex conjugate pairs, one can immediately satisfy (\ref{E4}) non-trivially, just as required. Conversely, if the secular equation that determines the energy eigenvalues is real, then $H$ and $H^{\dagger}$ will possess the same set of eigenvalues and thus be related by an isospectral similarity transform. Reality of the secular equation is thus both necessary and sufficient for the existence of a $V$ operator that obeys (\ref{E6}). Reality of the secular equation is thus both necessary and sufficient for unitary time evolution.

The reality requirement on a secular equation is a condition that is encountered in $PT$-symmetric theories ($P$ and $T$ denote parity and time reversal). Specifically, it was noted in \cite{R2} that if a Hamiltonian is $PT$ symmetric, the secular equation would be real. Moreover, it is not specifically $PT$ invariance itself that is required for the proof. All one needs is that $H$ be invariant under the action of the product of a discrete linear operator and a discrete anti-linear operator. For a time-independent $H$, time reversal reduces to complex conjugation $K$ times a linear operator that acts in the space of the eigenvectors of $H$, with $PT$ then acting as $AK$ where $A$ is a linear operator. The essence of the proof of the reality of the secular equation is to note that if a Hamiltonian commutes with $AK$, the secular equation is then of the form ${\rm det}(H-\lambda I)={\rm det}(AKHKA^{-1}-\lambda I)={\rm det}(KHK-\lambda I)={\rm det}(H^{*}-\lambda I)=0$, with reality of the secular equation immediately following. Discussion of the reality of secular equations dates back to Wigner \cite{R3}, who considered invariance of a Hamiltonian under time reversal alone. Modern interest in the reality of secular equations was triggered by the surprising discovery that the non-Hermitian Hamiltonian $H=p^2+ix^3$ had a completely real energy eigenspectrum \cite{R4,R5,R5a}, an outcome that was traced in \cite{R5,R5a} to the fact that while not time reversal invariant, $H$ was nonetheless invariant under the $PT$ product as $T$ transforms $i$ into $-i$ while $P$ transforms $x$ into $-x$. The utility of $PT$ invariance stems from the fact that complex potentials that are not $T$ invariant might still be invariant with respect to a $PT$ product where $P$ is an appropriate linear operator, and thus yield a real secular equation. 

As noted above, in \cite{R2} it was shown that if a Hamiltonian is $PT$ symmetric then its secular equation is real. More recently, the converse has been proven \cite{R6}, namely that if a Hamiltonian has a secular equation that is real, then one can always find appropriate $P$ and $T$ operators under which the Hamiltonian is $PT$ symmetric. Consequently, $PT$ symmetry of a Hamiltonian is both necessary and sufficient for reality of its secular equation, and thus $PT$ symmetry of a Hamiltonian is both necessary and sufficient for unitary time evolution.\footnote{This result has also been noted in \cite{R7b}.} The power of this result is that one can determine whether or not a Hamiltonian can generate unitary time evolution entirely by examining its symmetry structure under the action of the product of a discrete linear operator and a discrete anti-linear operator.\footnote{ In general in regard to issues involving Hermiticity, we should note the relation $H=H^{\dagger}$ has no invariant meaning. The relation $H_{ij}=H_{ji}^*$ is a basis-dependent statement that does not hold for the similarity transformed operator $H^{\prime}=SHS^{-1}$, since even if $H=H^{\dagger}$, then $H^{\prime \dagger}=S^{-1 \dagger}H^{\dagger}S^{\dagger}=S^{-1 \dagger}HS^{\dagger}=S^{-1 \dagger}S^{-1}H^{\prime}SS^{\dagger}$, with  $H^{\prime \dagger}$ only being equal to  $H^{\prime}$ if $S$ is unitary. Even though the statement that $H=H^{\dagger}$ is not an invariant one, the statement that the eigenvalues of an operator are all real and its eigenvectors are complete is an intrinsic, basis-independent, property of the operator itself. The eigenvectors are complete even if they are in a non-orthogonal basis, with their completeness entailing that one can find a basis in which they are orthogonal.  In this paper centrality will be  given to statements that are preserved by similarity transforms, statements such as $P^2=1$, $[PT,H]=0$ that involve products of operators.}

Within the family of Hamiltonians that have real secular equations there are three particular cases of interest  \cite{R6}: (i) the energies are all real and the energy eigenvectors are complete, (ii) the energy eigenvectors are complete but the energies include one or more complex conjugate pairs, and (iii) the energies are real but the energy eigenvectors do not form a complete set and the Hamiltonian is in non-diagonalizable, Jordan-block form. A possible fourth case, namely that the energy eigenvectors are not complete and the energies include one or more complex conjugate pairs, is not allowed for any individual $PT$-symmetric Jordan-block Hamiltonian since the energies of a Jordan-block matrix are all equal (they all share a single common eigenvector) and could thus not contain any complex conjugate pairs.\footnote{Even though no individual Jordan-block Hamiltonian could possess complex conjugate pairs of eigenvalues, a possible hybrid Jordan-block case was also noted in \cite{R6}, namely Jordan-block Hamiltonians that themselves come in pairs, one with all eigenvalues equal to some common $E$ and the other with all eigenvalues equal to a common $E^*$, and with $PT$ symmetry taking each of the two Jordan-block Hamiltonians in the pair into the other. Since the treatment of such hybrid cases follows that of individual $PT$-symmetric Jordan-block Hamiltonians given below in Sec. (5) we do not discuss it further here.} Moreover, in \cite{R6} criteria were also provided for determining to which of these three cases of interest any given $PT$-symmetric Hamiltonian belongs. Specifically, as noted above, it is very convenient \cite{R1} to introduce a linear operator, the $C$ operator, that obeys the two conditions: $[C,H]=0$, $C^2=I$.\footnote{While explicit construction of this $C$ operator is in general quite difficult, its construction can be greatly simplified if the Hamiltonian has specific properties. For instance, consider the case in which a $PT$-symmetric Hamiltonian is symmetric under transposition in some specific  basis (to thus obey $H=H^{\rm TR}$, $H^{\dagger}=H^*$ in that basis), and transforms as $THT=H^*$ under a time reversal operator $T$ that obeys $T^2=I$. Then in the basis a parity operator $P$ that obeys $P^2=I$ effects $PHP=H^{\dagger}$, a relation that often appears in $PT$ studies. With $V$ effecting $VHV^{-1}=H^{\dagger}$, then in a basis in which $PHP=H^{\dagger}$ we obtain $[PV,H]=0$. If in addition $P$ and $V$ are such that $PV^{-1}P=V$, the operator $C$ can be written as $C=V^{-1}P=PV$ and will then obey both $[C,H]=0$ and $C^2=I$.} In \cite{R6} it was shown that if every $C$ that non-trivially obeys these two conditions commutes with $PT$ then we are in case (i), that if there exists at least one $C$ that does not commute with $PT$ then we are in case (ii), and if there is no $C$ that obeys these two conditions at all then we are in case (iii). Thus as well as ascertain the unitarity structure of a Hamiltonian via a symmetry condition (i.e. whether or not $H$ commutes with $PT$), we can even determine which realization of the energy spectrum we are in by examining the structure of a commutator (viz.  that of $C$ with $PT$).  

Moreover, in \cite{R6} it was shown that in the event of $PT$ symmetry a $C$ operator that non-trivially obeys $[C,H]=0$, $C^2=I$ will in fact always exist if the eigenspectrum of the Hamiltonian is complete. Specifically, since one can always diagonalize any such Hamiltonian via a similarity transformation, any candidate $C$ operator that is to commute with the Hamiltonian would be diagonalized in the same basis. Then, since the eigenvalues of a diagonalizable $PT$-symmetric Hamiltonian are either real or appear in complex conjugate pairs, one only needs to use properties of two-dimensional matrices to establish that one can always find a non-trivial $C$ operator that obeys $C^2=I$ in the diagonal basis. Consequently, for diagonalizable $PT$-symmetric Hamiltonians there will always exist a non-trivial  $C$ operator that obeys $[C,H]=0$, $C^2=I$ in any basis. 

Now in practice it may not always be possible to construct such a $C$ operator in closed form. Nonetheless, the analysis of \cite{R6} shows that in principle such a $C$ operator will always exist. In completely solvable models the $C$ operator can be constructed in closed form, and in Secs. (3) and (4) we present a simple example. In this example we are able to vary parameters in the Hamiltonian continuously and go from a phase in which all energies are real and complete, to a phase in which all energy eigenvalues are real but the Hamiltonian is Jordan-block, and then continue on to a phase in which the energy eigenvalues appear in complex conjugate pairs and are again complete.  As we vary parameters the similarity transformation needed to diagonalize the Hamiltonian will become singular in the non-diagonalizable Jordan-block phase. As we will see, in the Jordan-block phase, the operator $C$ will become singular too.  Singularities in the $C$ operator thus signal the transitions between the different realizations of $PT$ symmetry as parameters are varied.  Having  now discussed some general aspects of $PT$ symmetry, we turn now to an examination of how the unitarity of time evolution works in each of the various realizations.

\section{Unitarity when energies are all real and complete}
\label{S3}

In the event that the energy spectrum of a $PT$-invariant Hamiltonian is both real and complete, there must exist a similarity transform $B$ that brings $H$ to a Hermitian form according to 
\begin{equation}
BHB^{-1}=\hat{H}
\label{E8a}
\end{equation}
where $\hat{H}$  is Hermitian. Then since $\hat{H}=\hat{H}^{\dagger}$ we can set
\begin{equation}
B^{\dagger}BHB^{-1}(B^{\dagger})^{-1}=H^{\dagger}.
\label{E8b}
\end{equation}
Thus in the event the energy spectrum of $H$ is both real and complete, one can always find an operator  $V=B^{\dagger}B$ that will effect $VHV^{-1}=H^{\dagger}$. Such a $V$ is not only automatically Hermitian, all of its eigenvalues are positive, i.e. $V$ is a positive operator.\footnote{Since $V$ is Hermitian it can be diagonalized by a unitary transformation according to $DVD^{\dagger}=V_{\rm D}$. Under this transformation  $B^{\dagger}B$ transforms into  $E^{\dagger}E$ where $E=DBD^{\dagger}$. For any diagonal component of $V_{\rm D}$ we can thus set $V^{\rm D}_{ii}=\sum_{k} E^{\dagger}_{ik}E_{ki}=\sum_{k}E^*_{ki}E_{ki}$, with all the eigenvalues of $V$  thus being positive.} Moreover, since the eigenvalues of $V_{\rm D}$ are positive, we are able to construct a diagonal operator  $G_{\rm D}$ according to $V_{\rm D}=G_{\rm D}^2$ where the eigenvalues of $G_{\rm D}$ are the positive square roots of the eigenvalues of $V_{\rm D}$, with the operator $G_{\rm D}$ thus also being both positive and Hermitian. On defining $G=D^{\dagger}G_{\rm D}D$, we thus see that we can set $V=G^2$ where $G$ is both Hermitian and a positive operator.\footnote{In the $PT$ literature the positive and Hermitian operators $G$ and $V$ are often written in the form $G=e^{-Q/2}$, $V=e^{-Q}$ where $Q$ is a Hermitian operator.}  (Analogous results have been reported in \cite{R1,R6,R7b,R7,R7a,R7c}.) As noted in \cite{R1,R7b,R8,R8a,R8b,R8d}, the utility of the operator $G$ is that since $G$ is Hermitian the operator $\tilde{H}=GHG^{-1}$ is Hermitian.\footnote{$(GHG^{-1})^{\dagger}=G^{-1}G^2HG^{-2}G=GHG^{-1}$.} Since we have constructed two separate operators that are Hermitian, viz. $BHB^{-1}=\hat{H}$ and $GHG^{-1}=\tilde{H}$, these two operators must be unitarily equivalent. With $\hat{H}=BG^{-1}\tilde{H}GB^{-1}$, with $B$ and $G$ being related via $B^{\dagger}B=G^2$, and with $G$ being Hermitian, we obtain $(GB^{-1})^{\dagger}=BG^{-2}G=BG^{-1}=(GB^{-1})^{-1}$, with the operator $GB^{-1}$ thus being unitary just as required.\footnote{Strictly speaking, the restriction to positive $G$ is not mandatory, since if we constructed a $G^{\prime}_{\rm D}$ whose eigenvalues are the negative square roots of  the eigenvalues of $V$, through the use of $G^{\prime}=D^{\dagger}G^{\prime}_{\rm D}D$ we would recover the same $\tilde{H}$ that we get using $G$. We could thus set $V=(G^{\prime})^2$ and use a Hermitian $G^{\prime}$ whose eigenvalues are all negative.} In an Appendix we demonstrate the above analysis in a simple solvable case.\footnote{In making similarity transformations in the presence of  $PT$ symmetry, we should note that we need to take into account the fact that $T$ is anti-linear. Specifically, if we set $P=\pi$, $T=\tau K =K \tau^*$ where $\pi$ and $\tau$ are linear operators and $K$ denotes complex conjugation, and require that $P^2=I$, $T^2=I$, $[P,T]=0$, we obtain $\pi^2=I$, $\tau\tau^*=I$, $\pi\tau=\tau\pi^*$. If we now make a similarity transform $SPS^{-1}=P^{\prime}$, $STS^{-1}=T^{\prime}$ and set $P^{\prime}=\pi^{\prime}$, $T^{\prime}=\tau^{\prime}K$, then  with $\pi^{\prime}=S\pi S^{-1}$, $\tau^{\prime}=S\tau (S^{-1})^*$ we obtain $P^{\prime 2}=I$, $T^{\prime 2}=I$, $[P^{\prime},T^{\prime}]=0$, and $\pi^{\prime 2}=I$, $\tau^{\prime} \tau^{{\prime}*}=I$, $\pi^{\prime}\tau^{\prime}=\tau^{\prime}\pi^{{\prime}*}$. If we transform a Hamiltonian $H$ obeying $H=PTHTP=\pi\tau H^*\tau^*\pi$, we find that $H^{\prime}=SHS^{-1}=P^{\prime}T^{\prime}H^{\prime}T^{\prime}P^{\prime}=\pi^{\prime}\tau^{\prime} H^{{\prime} *}\tau^{{\prime}*}\pi^{\prime}$, with $PT$ symmetry being maintained.}

In the event that the energy spectrum of a $PT$-invariant Hamiltonian contains complex conjugate pairs, or in the event that the energy spectrum is not in fact complete, one can still find a $V$ that effects $VHV^{-1}=H^{\dagger}$. However, in neither of these two cases can the relevant $V$ operator  be written in a  $V=B^{\dagger}B$ form. We shall discuss these latter two cases further below, but for the moment we shall concentrate on the case where the energy spectrum is both real and complete

In the particular case where the energy spectrum is both real and complete we can use the operators $G$ and $V$ to construct an appropriate norm. Specifically,  we note that if for $\tilde{H}$ we introduce a standard set of right-eigenstates $|\tilde{n}^i\rangle$, the right-eigenstates of $H$ are thus related to those of $\tilde{H}=GHG^{-1}$ by the mapping $|n_i\rangle=G^{-1}|\tilde{n}_i\rangle$. Since $G$ is Hermitian, and since it therefore effects a mapping that is not unitary, the states  $\langle n_i|=\langle \tilde{n}_i|G^{-1}$ are not left-eigenstates of $H$. Rather, it is the states $\langle n_i|G^2=\langle n_i|V$ that are the left-eigenstates. Moreover, since the eigenstates of the Hermitian $\tilde{H}$ obey the standard time-independent Dirac orthonormality condition $\langle \tilde{n}^i|\tilde{n}^j\rangle=\delta_{i,j}$, then by the mapping the eigenstates of $H$ obey the time-independent condition $\langle n^i|V|n^j\rangle=\delta_{i,j}$. With recognize the inner product  $\langle n^i|V|n^j\rangle$ as being none other than the time-independent $V$-operator norm introduced above.\footnote{With $C$ commuting with $PT$ when all energies are real, we can take all the energy eigenstates to be eigenstates of $CPT$ with eigenvalue one. In cases where we are additionally able to identify $C$ as $C=V^{-1}P$, and thus identify $C=PV$ and $C^{\dagger}=PV^{-1}$ (which requires that $P$, and not just $V$, be Hermitian), we can write the norm as $(({CPT}u(t))^{\dagger},Vv(t))=(u^{\rm TR}(-t),P^{\dagger}Pv(t))=(u^{\rm TR}(-t),v(t))=(({CPT}u(t))^{\rm TR},v(t))$, to thus recover the norm used in \cite{R1}. Now in $PT$ studies it is often the case that the operator $V$ is non-local (by, for instance, involving an exponential of the momentum operator). However, for those cases in which one can set $C=PV$, then when taken between eigenstates of the Hamiltonian and thus between eigenstates of $C$ as well, matrix elements of the form $\langle 1|V|2\rangle$ can be written as $\pm \langle 1|P|2\rangle$, since $C^2=I$. In consequence, even if $V$ is non-local, matrix elements of the form $\langle n^i|V|n^j\rangle=\langle \tilde{n}^i|\tilde{n}^j\rangle=\delta_{i,j}$ possess the standard local positive-definite probability interpretation that we described above.} 

The essence of the above analysis is the realization that when a Hamiltonian $H$ has an energy spectrum that is both real and complete, the Hamiltonian is either already Hermitian or can be brought to a Hermitian form $\tilde{H}$ by a similarity transformation. However, if a Hamiltonian $H$ is not Hermitian the needed similarity transformation will not be unitary and the basis states of $H$ will not be orthonormal with respect to the Dirac inner product. Rather, they will form a skew basis, with it being the similarity transform that puts the basis into an orthonormal form. $PT$-symmetric Hamiltonians that are not Hermitian but nonetheless have a real and complete energy spectrum are thus Hermitian in disguise. Consequently, they are unitary in disguise too, with it being the $V$-operator norm that is the appropriate norm for the Hamiltonian.

It is instructive to recast the above discussion in the language of the $S$-matrix. For the above $|\tilde{n}_i\rangle$ basis first, we can define complete sets of in and out states that obey the standard closure relations $\sum_i|\tilde{n}^i_{in}\rangle\langle \tilde{n}_{in}^i|=I$ and  $\sum_i|\tilde{n}^i_{out}\rangle\langle \tilde{n}_{out}^i|=I$. Given these relations one can show that the scattering operator defined by $\tilde{S}=\sum_i|\tilde{n}_{in}^i\rangle\langle \tilde{n}_{out}^i|$ and thus by $\tilde{S}^{\dagger}=\sum_i|\tilde{n}_{out}^i\rangle\langle \tilde{n}_{in}^i|$ immediately obeys the unitarity relation  $\tilde{S}\tilde{S}^{\dagger}=I$. 

Since the in and out states of $\tilde{H}$ can be mapped into the in and out states of $H$ by the $G$ mapping, an $S$-matrix operator in the $|n_i\rangle$ basis can thus be defined via the mapping as $\tilde{S}=GSG^{-1}$, with the adjoint $S^{\dagger}$ being given via $\tilde{S}^{\dagger}=G^{-1}S^{\dagger}G$ since the operator $G$ is Hermitian. As constructed, the operator $S$ can be written as $S=G^{-1}\sum_i|\tilde{n}_{in}^i\rangle\langle \tilde{n}_{out}^i|G$, and from the mapping thus takes the form $S=\sum_i|n_{in}^i\rangle\langle n_{out}^i|G^2$ in the $|n_i\rangle$ basis. In addition, the $\tilde{S}\tilde{S}^{\dagger}=I$ relation obliges $S$ to obey $GSG^{-1}G^{-1}S^{\dagger}G=I$, i.e. to obey 
\begin{equation}
SG^{-2}S^{\dagger}G^2=I,\qquad G^2SG^{-2}S^{\dagger}=I.
\label{E9}
\end{equation}
It is thus (\ref{E9}) that is the form that unitarity of a scattering process takes in a skew basis.\footnote{Because the only role of $G$ is to convert a skew basis into an orthogonal one according to $G|n_i\rangle=|\tilde{n}_i\rangle$, any possible non-locality in $G$ or in $G^2=V$ would not prevent one from obtaining the standard $\tilde{S}\tilde{S}^{\dagger}=I$. Thus the only impediment to defining an S-matrix at all would be the standard one, namely any non-locality of the interactions that are to transform in-states into out-states, with any possible non-locality in $G$ or $V$ not being a cause for concern.}

While we have derived (\ref{E9}) by transforming to the Hermitian basis and then transforming back, it  is possible to derive (\ref{E9}) directly in the $|n_i\rangle$ basis without needing to make any reference to the tilde basis at all. Moreover, there is a more general derivation that does not even require us to restrict to the Hermitian $V$ associated with the case in which all energy eigenvalues are real. We can allow for energies that appear in complex conjugate pairs as well, but for either real or complex energies we must still require that the energy eigenstates be complete. Specifically, we note that for energy eigenstates, be they real or complex, it follows from  (\ref{E5}) and (\ref{E6}) that the time independence of  $\langle R_j(t)|V|R_i(t)\rangle$  entails that states whose energies do not obey (\ref{E4}) have to obey $\langle R_j(t=0)|V|R_i(t=0)\rangle=0$. Consequently, such states are orthogonal with respect to the $V$-operator norm, and on normalizing the states appropriately, for all energy eigenstates we can thus set $\langle R_j(t)|V|R_i(t)\rangle=\delta_{i,j}$. With this orthonormality relation we obtain orthonormality and closure relations for the in and out states of the form
\begin{eqnarray}
&&\langle n_{in}^i|V|n_{in}^j\rangle=\langle n_{in}^j|V^{\dagger}|n_{in}^i\rangle=\delta_{i,j},\qquad \langle n_{out}^i|V|n_{out}^j\rangle=\langle n_{out}^j|V^{\dagger}|n_{out}^i\rangle=\delta_{i,j},
\nonumber\\
&&\sum_i|n^i_{in}\rangle\langle n_{in}^i|V=
\sum_iV^{\dagger}|n^i_{in}\rangle\langle n_{in}^i|=I,~~
\sum_i|n^i_{out}\rangle\langle n_{out}^i|V= \sum_iV^{\dagger}|n^i_{out}\rangle\langle n_{out}^i|=I.
\label{E10}
\end{eqnarray}
If we now define an operator $S=\sum_i|n_{in}^i\rangle\langle n_{out}^i|V$, from (\ref{E10}) we obtain
\begin{equation}
SV^{-1}S^{\dagger}V=\sum_i|n_{in}^i\rangle\langle n_{out}^i|VV^{-1}\sum_jV^{\dagger}|n_{out}^j\rangle\langle n_{in}^j|V=\sum_i|n_{in}^i\rangle\langle n_{in}^i|V=I.
\label{E11}
\end{equation}
Consequently, for $PT$-invariant Hamiltonians unitarity of the scattering process thus takes the equivalent forms
\begin{equation}
SV^{-1}S^{\dagger}V=I,\qquad VSV^{-1}S^{\dagger}=I.
\label{E12}
\end{equation}

As a simple check on (\ref{E12}), we note that if $S$ is written as the evolution operator $U=e^{-iHt}$, then because $H^{\dagger}=VHV^{-1}$, it follows that $VUV^{-1}$ is given as $e^{-iH^{\dagger}t}$. Then with $U^{\dagger}=e^{+iH^{\dagger}t}$, (\ref{E12}) immediately follows.\footnote{Unitary time evolution in the form $UV^{-1}U^{\dagger}V=I$ given in (\ref{E8p}) has also been discussed in \cite{R8e}. An equivalent formulation of unitarity in the $PT$ case may be found in \cite{R8ee}, where it is noted that one can write $SS^{PT}=I$ where $S^{PT}=(PT)S(PT)^{-1}$ is the $PT$ conjugate of $S$.  This relation follows since $(PT)e^{-iHt}(PT)^{-1}=e^{+iHt}$ if $[H,PT]=0$. Since $C$ commutes with $H$, one can also set $SS^{CPT}=I$ where $S^{CPT}=(CPT)S(CPT)^{-1}$ is the $CPT$ conjugate of $S$.} The relation given in (\ref{E12}) is thus the unitarity relation we want. While we shall discuss the complex energy case in more detail below, as we had already noted, our derivation of (\ref{E12}) holds even if some energies are complex. The issue here is not whether or not we can prepare stable in and out states with them, but whether or not we need them for the closure relations given in (\ref{E10}). In other words, suppose we have a $PT$-invariant Hamiltonian with $N$ real energy eigenvalues and $M$ complex pairs. For such a situation, unitarity in the form given in (\ref{E12}) will only follow if in the closure relations in (\ref{E10}) we sum over $N+2M$ states (i.e. over $N+2M$ channels), and not over $N$ states alone. For an unstable  state that decays into some decay products, it is not possible to construct a scattering  experiment in which the decay products can be recombined into the original unstable state. Nonetheless the states associated with the decay product configurations still need to be included in the closure relations.

As well as provide a formal derivation of our results, it is  instructive to see how our ideas work in a simple model. Since we can always diagonalize any Hamiltonian that possesses a complete set of eigenstates, and since complex conjugate pairs would form a two-dimensional sub-block in the diagonal basis, it will suffice to discuss a two-dimensional model. We thus follow \cite{R8f} and introduce the Hamiltonian (the parameters $r$, $s$ and $\theta$ are real)
\begin{equation}
H=r\cos\theta\sigma_0+ir\sin\theta \sigma_3+s\sigma_1=\left(\matrix{r\cos\theta+ir\sin\theta&s\cr s&r\cos\theta-ir\sin\theta\cr}\right),
\label{E13}
\end{equation}
with $H$ being symmetric under transposition. While not Hermitian, this $H$ is $PT$ invariant under $P=\sigma_1$, $T=K$ and thus has a real secular equation with energies being given by $E_{\pm}=r\cos\theta \pm (s^2-r^2\sin^2\theta)^{1/2}$ for any choice of values of the parameters  $r$, $s$ and $\theta$. In the region where $s^2-r^2\sin^2\theta$ is positive the two energy eigenvalues are real and distinct. In this region we can define Hermitian operators $G$ and $V$ according to
\begin{equation}
G^{\pm 1}=\left({1+\sin\alpha\over 2\sin\alpha}\right)^{1/2}\sigma_0\pm \sigma_2\left({1-\sin\alpha\over 2\sin\alpha}\right)^{1/2},\qquad 
G^{\pm 2}=V^{\pm 1}={1 \over \sin\alpha}\sigma_0\pm\sigma_2 {\cos\alpha \over \sin\alpha},
\label{E14}
\end{equation}
where $\sin\alpha=+(s^2-r^2\sin^2\theta)^{1/2}/s$, $\cos\alpha=r\sin\theta/s$.\footnote{The eigenvalues of $G$ are  $[(1+\sin\alpha)/2\sin\alpha]^{1/2}\pm [(1-\sin\alpha)/2\sin\alpha]^{1/2}$ while the eigenvalues of $G^2$ are $(1\pm \cos\alpha)/\sin\alpha$, and all are positive in the domain where $s>r\sin\theta$.} Using the $G$ operator we construct
\begin{equation}
\tilde{H}=GHG^{-1}=r\cos\theta\sigma_0+\sigma_1(s^2-r^2\sin^2\theta)^{1/2},
\label{E15}
\end{equation}
and confirm that $\tilde{H}$ is Hermitian when $s^2-r^2\sin^2\theta$ is positive. Similarly we construct $VHV^{-1}$ and confirm that it is equal to $H^{\dagger}$. Finally, we construct an operator $C=V^{-1}P$, viz.
\begin{equation}
C={1 \over \sin\alpha}\left(\sigma_1+i\cos\alpha\sigma_3\right),
\label{E16}
\end{equation}
and confirm that it not only obeys $[C,H]=0$ and $C^2=I$, it also obeys ${[C,PT]}=0$, just as it must \cite{R6} since the energies are real.\footnote{As constructed, the eigenvalues of the $C$ operator given in (\ref{E16}) are $+1$  and $-1$. With the eigenvalues of the $P$=$\sigma_1$ operator also being $+1$  and $-1$ in this case, we see that even though neither  $C$ nor $P$ is a positive operator their product ${CP}=V^{-1}$ is.}

In the region where $s^2-r^2\sin^2\theta$ is positive, eigenvector solutions that obey $idu_{\pm}/dt=Hu_{\pm}=E_{\pm}u_{\pm}$ are given as 
\begin{equation}
u_+={e^{-i(r\cos\theta+\mu)t} e^{i\pi/4}\over (2\sin\alpha)^{1/2}}\left(\matrix{e^{-i\alpha/2}\cr -ie^{i\alpha/2}\cr}\right),\qquad
u_-={e^{-i(r\cos\theta-\mu)t} e^{i\pi/4}\over (2\sin\alpha)^{1/2}}\left(\matrix{ie^{i\alpha/2}\cr e^{-i\alpha/2}\cr}\right).
\label{E17}
\end{equation}
where $\mu=+(s^2-r^2\sin^2\theta)^{1/2}$. For the states in (\ref{E17}) the $V$ norms obey the time-independent and manifestly unitary orthonormality and closure relations 
\begin{equation}
u_{\pm}^{\dagger}Vu_{\pm}=+1,\qquad u_{\pm}^{\dagger}Vu_{\mp}=0,\qquad u_{+}u^{\dagger}_{+}V+u_{-}u^{\dagger}_{-}V=I,
\label{E18}
\end{equation}
just as required. It is thus the $V=G^2$ norm that it is the appropriate one for the problem. 

Finally, as we discuss in detail below, we find that when we set $s^2-r^2\sin^2\theta=0$, the $E_+$ and $E_-$  energy eigenvalues become equal and the Hamiltonian becomes a non-diagonalizable Jordan-block Hamiltonian. Consequently, $G$, $V$ and $C$ must become singular at this point, just as can be seen above. Then, if we continue on to $s^2-r^2\sin^2\theta<0$ the eigenvalues become complex and, as we now show,  $G$, $V$ and $C$ become non-singular again.

\section{Unitarity when energies are not all real but complete}
\label{S4}

To determine what happens when energies are complex, it is instructive to study the two-dimensional model in the regime where $s^2-r^2\sin^2\theta$ is negative. Now the energies are given by the complex conjugate pair $E_{\pm}=r\cos\theta\pm i\nu$ where $\nu=+(r^2\sin^2\theta-s^2)^{1/2}$.
In this region we can define now non-Hermitian  operators $G$ and $V$ according to an appropriate continuation of (\ref{E14}), viz. 
\begin{equation}
G^{\pm 1}=\left({1+i\sinh\beta\over 2i\sinh\beta}\right)^{1/2}\sigma_0\pm \sigma_2\left({1-i\sinh\beta\over 2i\sinh\beta}\right)^{1/2},\qquad G^{\pm 2}=V^{\pm 1}={1 \over i\sinh\beta}\sigma_0\pm\sigma_2 {\cosh\beta \over i\sinh \beta},
\label{E19}
\end{equation}
where $\sinh\beta=+(r^2\sin^2\theta-s^2)^{1/2}/s$, $\cosh\beta=r\sin\theta/s$.\footnote{The eigenvalues of $G$ are  $[(1+i\sinh\beta)/2i\sinh\beta]^{1/2}\pm [(1-i\sinh\beta)/2i\sinh\beta]^{1/2}$ while the eigenvalues of $G^2$ are $(1\pm \cosh\beta)/i\sinh\beta$, and neither is a positive operator in the domain where $r\sin\theta>s$.}  Using this $G$ operator we construct
\begin{equation}
\tilde{H}=GHG^{-1}=r\cos\theta\sigma_0+i\sigma_1(r^2\sin^2\theta-s^2)^{1/2},
\label{E20}
\end{equation}
and see that $\tilde{H}$ is not Hermitian when $s^2-r^2\sin^2\theta$ is negative. Despite this,  we construct $VHV^{-1}$ with this non-Hermitian $V$ and confirm that it nonetheless is equal to $H^{\dagger}$, just as it must be, with the relevant $V$ thus not being Hermitian this time and thus not being writable in the form $V=B^{\dagger}B$ for any choice of $B$. Finally, we construct an operator $C=V^{-1}P$ with $P$ still equal to $\sigma_1$ , viz.
\begin{equation}
C={1 \over i\sinh\beta}\left(\sigma_1+i\cosh\beta\sigma_3\right)
\label{E21}
\end{equation}
and confirm that it while it obeys $[C,H]=0$ and $C^2=I$, because of its overall factor of $i$ it does not obey ${[C,PT]}=0$, as must be the case \cite{R6} since the energies are not real. As required, the $C$ operator given in (\ref{E21}) is singular in the Jordan-block limit where $r^2\sin^2\theta-s^2=0$.

In the region where $s^2-r^2\sin^2\theta$ is negative, eigenvector solutions that obey $idu_{\pm}/dt=Hu_{\pm}=E_{\pm}u_{\pm}$ can be constructed by setting $\mu=i\nu$, $\alpha=i\beta$ in (\ref{E17}), and are given as 
\begin{equation}
u_+={e^{-ir\cos\theta t+\nu t} \over (2\sinh\beta)^{1/2}}\left(\matrix{e^{\beta/2}\cr -ie^{-\beta/2}\cr}\right),\quad
u_-={e^{-ir\cos\theta t-\nu t} \over (2\sinh\beta)^{1/2}}\left(\matrix{ie^{-\beta/2}\cr e^{\beta/2}\cr}\right).
\label{E22}
\end{equation}
For the states in (\ref{E22}) the $V$ norms obey the time-independent orthogonality and closure relations 
\begin{eqnarray}
&&u_{\pm}^{\dagger}Vu_{\pm}=0,\qquad u_{-}^{\dagger}Vu_{+}=+ 1,\qquad u_{+}^{\dagger}Vu_{-}=- 1,
\nonumber\\
&&u_{+}u^{\dagger}_{-}V-u_{-}u^{\dagger}_{+}V=I.
\label{E23}
\end{eqnarray}
With the adjoints of the states in (\ref{E22}) having time dependences of the form $u_{+}^{\dagger}\sim e^{ir\cos\theta t+\nu t}$,  $u_{-}^{\dagger}\sim e^{ir\cos\theta t-\nu t}$, we see that in (\ref{E23}) those overlaps that are zero have to vanish through the $V$ matrix structure of the overlaps since the time dependences in $u_{\pm}^{\dagger}Vu_{\pm}$ do not cancel each other once $\nu$ is non-zero. For the non-zero overlaps the time dependences in $u_{\mp}^{\dagger}Vu_{\pm}$ do precisely cancel each other, just as required by (\ref{E4}). With the $u_+$ and $u_-$ states in (\ref{E22}) obeying $PTu_+(t)=u_-(-t)$, $PTu_-(t)=u_+(-t)$ (for time-dependent functions $T$ effects $u(t)\rightarrow u^*(-t)$), the non-vanishing overlaps in (\ref{E23}) represent transitions between energy eigenstates and their $PT$ partners.\footnote{This transition structure also holds when all energy eigenvalues are real. With the overall phases of the states in (\ref{E17}) being chosen so that the  eigenvalues of the anti-linear $PT$ operator are equal to $+1$ and $-1$ (so that $PTu_{\pm}(t)=\pm u_{\pm}(-t)$, $Cu_{\pm}(t)=\pm u_{\pm}(t)$, ${CP}Tu_{\pm}(t)=u_{\pm}(-t)$), the non-zero overlaps are between states and themselves as they are their own $PT$ partners. (This $PT$ structure of energy eigenstates is generic, since, as noted in  \cite{R1},  for real energies  the energy eigenstates are eigenstates of $PT$, and for complex conjugate energies the energy eigenstates transform into each other under $PT$.)}

The most notable aspect of the overlaps is that the $u_{+}^{\dagger}Vu_{-}$ overlap is negative. However, this does not herald the presence of a negative metric since the overlap is a transition matrix element between different states and not the overlap of a state with its own adjoint. Moreover, rather than being an indicator of a possible loss of unitarity,  this negative sign is actually needed for unitarity. Specifically, if we construct a propagator or resolvent operator for the above Hamiltonian, it will have the form of matrix elements of the states divided by energy denominators: 
\begin{equation}
D(E)=\frac{u_{-}^{\dagger}Vu_{+}}{E-(E_R-iE_I)}+\frac{u_{+}^{\dagger}Vu_{-}}{E-(E_R+iE_I)},
\label{E24}
\end{equation}
and thus evaluate to
\begin{equation}
D(E)=\frac{1}{E-(E_R-iE_I)}-\frac{1}{E-(E_R+iE_I)}=\frac{-2iE_I}{(E-E_R)^2+E_I^2}.
\label{E25}
\end{equation}
Other than the overall factor of $2$, the imaginary part of this propagator is the same as that of a standard Breit-Wigner
\begin{equation}
D_{\rm BW}(E)=\frac{1}{E-(E_R-iE_I)}=\frac{E-E_R-iE_I}{(E-E_R)^2+E_I^2}.
\label{E26}
\end{equation}
The negative sign of $u_{+}^{\dagger}Vu_{-}$ thus compensates for the positive sign of the imaginary part of the energy of the $u_{+}$ state,  with its contribution  to the imaginary part of $D(E)$ then being negative rather than positive.

In passing we note that an effect  similar to this  is observed in Lee-Wick electrodynamics \cite{R11,R11a,R11b}, where a pair of states with complex conjugate energies combine as in (\ref{E25}) to produce a unitary propagator. The difference between the approach here and that of Lee and Wick is that Lee and Wick attributed the needed minus sign in (\ref{E25}) to an intrinsic property of the Hilbert space by employing a preassigned indefinite metric in which $u_-^{\dagger}u_-$ (the overlap of $u_-$ with itself) is taken to be negative and the closure relation is of the generic, so-called Krein space, form $u_{+}u^{\dagger}_{+}-u_{-}u^{\dagger}_{-}=I$. In our work here we do not work with an indefinite metric, nor do we preassign the metric. Rather the $V$ inner product is determined by the theory itself, with the $V$ norm not being a universal norm but being one that depends each time on the particular form of the Hamiltonian, as it is defined as the operator that effects  $VHV^{-1}=H^{\dagger}$. (As noted in \cite{R11f}, the indefinite Krein space norm is equivalent to the $PT$ norm in which one builds scalar products out of states and their $PT$ conjugates, whereas the $V$ norm is equivalent to the $CPT$ norm in which one builds scalar products out of states and their $CPT$ conjugates with the $C$ operator being Hamiltonian dependent.) 

In fact, in general the key difference between the norms needed for Hermitian theories and $PT$ theories is that while the Hermitian case norm can be assigned a priori (c.f. the Dirac norm or its quasi-Hermitian negative metric Krein space generalization), for $PT$ theories each Hamiltonian determines its own $V$ norm. It was by taking advantage of this fact that Bender and Mannheim \cite{R11c,R11d,R11e} were able to show that theories based on the fourth-order propagator $D(k^2)=1/k^2-1/(k^2+M^2)$ were unitary theories that did not need to be formulated in an indefinite metric Hilbert space. (With the poles of this fourth-order propagator being both real and complete, the associated Hilbert space closure relation is of the generic form $u_{+}u^{\dagger}_{+}V+u_{-}u^{\dagger}_{-}V=I$ given in (\ref{E18}), with the relative minus sign in the fourth-order propagator being associated with a negative eigenvalue of a $C$ operator  that obeys $C=PV$ \cite{R11e}.) Moreover, by establishing that the fourth-order derivative conformal gravity theory is a $PT$ theory, conformal gravity is now able to emerge as a fully renormalizable and unitary theory of quantum gravity in four spacetime dimensions \cite{R11f,R11ga,R11g,R11h}. Some other examples in which  $PT$ symmetry has been used to remove Dirac norm ghosts and secure unitarity may be found in \cite{R11i} and \cite{R11j}.

While it is nice to see that $D(E)$ recovers the standard Breit-Wigner form as given above, our results actually go further than that as they provide some justification for the use of the standard $D_{\rm BW}(E)$ in the first place. Specifically, if one works with a general complex potential, i.e. with one for which (\ref{E4}) is not obeyed, one would not get unitary time evolution. If for instance we work with a complex potential that has just one energy eigenstate with wave function $\psi=e^{-iE_Rt+E_It}$, the probability would be given as  $\psi^*\psi=e^{-2E_It}$ and not be preserved in time, and would even grow in time if $E_I$ is negative. Hence, if we wish to describe absorption or decay processes in a time-preserving manner, we must augment this state with its complex conjugate partner and include the contribution of the partner to $D(E)$. Our approach is thus well-suited to a scattering process in which the in and out states long before and long after a collision are real eigenvalue eigenstates of a $PT$-invariant Hamiltonian $H_0$, with the states scattering through some short-range $PT$-invariant interaction $H_{\rm INT}$ for which $H_0+H_{\rm INT}$ has complex eigenvalues. In such a situation the interaction would generate two poles at $E_R\pm iE_I$ that would be on the same Riemann sheet and combine just as in (\ref{E25}).\footnote{It would be of interest to see if the factor of $2$ in (\ref{E25}) might lead to some observable consequence, perhaps in the determination of branching ratios.}

The fact that the two complex conjugate poles are on the same Riemann sheet has an interesting consequence: it enables the two poles to be located within the same contour in the complex energy plane rather than be  in different contours. In consequence, the theory will be causal. Specifically, with two complex conjugate poles having opposite imaginary parts, one of the poles would have to appear above the real energy axis and one below. Now when one only has to deal with poles with real energies (such as the $D(E)=1/E_++1/E_-$ propagator associated with  (\ref{E18})), to construct a causal, retarded propagator, one locates all the poles slightly below the real axis, with a contour integration then giving no pole contribution if one closes the contour in the upper-half plane. If one simply repeats this prescription when there is a pole in the upper-half plane, on closing the contour above the real axis one then gets a non-zero pole contribution and causality violations result. However, as discussed for instance in \cite{R13}, for complex conjugate pairs of poles one can use an alternate contour instead. Specifically, one can deform the contour to pass above any  pole in the upper-half plane. Then when one closes this deformed contour in the upper-half plane, one still gets no pole contribution and the propagator is causal, and when one closes the deformed contour in the lower-half plane, both complex poles contribute, just as in (\ref{E25}). For $PT$ theories then, to maintain causality, the prescription is to make the choice of contour dynamics dependent (just as is done with the $PT$-theory Hilbert space metric), so that as interactions cause poles to move off the real axis the contour must move along with them so that all the poles remain below it, a meaningful procedure if all the poles are on the same Riemann sheet.\footnote{It could be of interest to see if the causality problems that are met in Lee-Wick electrodynamics \cite{R11,R11a,R11b,R13} might originate in the use of a preassigned Krein space norm rather than the dynamically determined $V$ norm.}

\section{Unitarity when energies are all real but incomplete}
\label{S5}

When a Hamiltonian is diagonalizable,  one does not need to distinguish between the eigenvalue solutions to the secular equation ${\rm det}(H-\lambda I)=0$ and the eigenvalue solutions to $H\psi=E\psi$ since the two sets of solutions coincide. However,  when a Hamiltonian is not diagonalizable, the number of eigensolutions to $H\psi=E\psi$ is less than the number of eigenvalue solutions to ${\rm det}(H-\lambda I)=0$, and the spectrum of $H$ is incomplete. A typical example of a non-diagonalizable, Jordan-block, matrix is the two-dimensional matrix $M$
\begin{eqnarray}
M=\pmatrix{a&1\cr 0&a}.
\label{E27}
\end{eqnarray}
Its secular equation $|M-\lambda I|=0$ has two solutions for $\lambda$, both of which are equal to $a$ (and incidentally both solutions are real when $a$ is real even though an $M$ with real $a$ is not Hermitian), but  $M$ has only one right-eigenvector and equally only one left-eigenvector, viz.
\begin{eqnarray}
\pmatrix{a&1\cr 0&a}\pmatrix{1\cr 0}=a\pmatrix{1\cr 0},\qquad a\pmatrix{0 & 1}=\pmatrix{0 & 1}\pmatrix{a&1\cr 0&a}.
\label{E28}
\end{eqnarray}

Despite the lack of completeness of its eigenstates, if the secular equation ${\rm det}(H-\lambda I)=0$ of a Hamiltonian is real, $H$ and its adjoint $H^{\dagger}$ will still be related by (\ref{E6}), i.e. still be related $VHV^{-1}=H^{\dagger}$ for some operator $V$. For the example of (\ref{E27}) the secular equation will be real when $a$ is real, and $V$ will then be given by $V=\sigma_1$.\footnote{For the 2-dimensional example given in (\ref{E13}), when we let $\alpha$ go to zero the two eigenvalues $E_+$ and $E_-$ both become become equal to $r\cos\theta$, and the two eigenvectors $u_+$ and $u_-$  in (\ref{E17}) collapse onto a single eigenvector, with the Hamiltonian becoming Jordan block in the limit. At the point where $\alpha$ becomes zero, both $V$ and $C$ become undefined, just as they must because $\alpha=0$ is the transition point between the phase where the energies are real and the phase where they are complex. Despite this, $H$ and $H^{\dagger}$ themselves (as well as $P$ and $T$) remain well-defined when $\alpha\rightarrow 0$ since one just sets $s=r\sin\theta$ in $H$. Despite the fact that both $V$ and $V^{-1}$ become undefined in the limit,  the product $VHV^{-1}$ remains well-defined,  and thus continues to be equal to $H^{\dagger}$.} More  generally, for a non-diagonalizable matrix of arbitrary dimensionality, Jordan showed that via a sequence of similarity transformations any such matrix can always be brought to the Jordan canonical form in which all of the elements on the diagonal are equal to each other, and the only non-zero off-diagonal elements are all real and all lie on the superdiagonal immediately above the diagonal. With the non-zero elements on the superdiagonal not contributing to the secular equation since all the other non-diagonal elements of the matrix are zero, the elements on the diagonal of a Jordan-block matrix are the eigenvalue solutions to the secular equation. Since all the eigenvalues of a Jordan-block matrix have to be equal to each other, the reality of the secular equation of a Jordan-block matrix requires that  all of its solutions be real, since any complex conjugate pair of solutions could not be equal to each other. Thus for any Jordan-block matrix whose secular equation is real, we see that all the elements of the matrix are real, with the Hermitian adjoint of the matrix thus being equal to its transpose. Since a transposition matrix always exists in any dimension (an explicit construction is given in \cite{R6}), the transposition matrix serves as $V$.  Thus for any Jordan-block matrix with a real secular equation, one can always find an operator $V$ that effects $VHV^{-1}=H^{\dagger}$. 

Moreover, in the Jordan-block case the converse also holds. Specifically  if one can find  a $V$ operator that effects $VHV^{-1}=H^{\dagger}$, the secular equation will be real since the invariance of a determinant under similarity transforms is insensitive to Jordan-block structures. Thus even in the Jordan-block case $PT$ invariance is necessary and sufficient for the existence of a $V$ operator that effects $VHV^{-1}=H^{\dagger}$.

With (\ref{E5}) holding for any states that obey (\ref{E1}) and (\ref{E2}), (\ref{E5}) will hold for non-stationary as well as stationary solutions to (\ref{E1}). Now in the Jordan-block case the lack of energy eigenstates is compensated for (see e.g. \cite{R11e}) by the presence of non-stationary solutions that typically behave as powers of $t$. For every missing stationary solution one finds a power-behaved in $t$ solution.\footnote{As the parameter $\mu$ becomes very small, the  $e^{- i\mu t}$ and $e^{+ i\mu t}$ phases of the two typical solutions given in (\ref{E17}) behave as $(1-i\mu t)$ and $(1+i\mu t)$, with the combination $(u_{+}-u_{-})/\mu$ limiting to a form that is linear in $t$.} The set of all stationary plus non-stationary solutions combined is thus complete.\footnote{Since non-diagonalizable Hamiltonians can be constructed as limits of diagonalizable ones, the completeness of the eigenstates of a diagonalizable Hamiltonian translates into the completeness of the stationary plus non-stationary solutions to the non-diagonalizable one, with the counting being as given in \cite{R11e}.} Since the stationary plus non-stationary solutions are complete, we can use (\ref{E5}) to infer that in the Jordan-block case time evolution of wave packets will be unitary if and only if one can find a $V$ operator that effects $VHV^{-1}=H^{\dagger}$. Thus, as noted in \cite{R11e}, even though the non-stationary solutions are power-behaved in $t$, packets containing them and the stationary solutions will still evolve with a probability that is preserved in time.\footnote{As noted above, for a diagonalizable $PT$-symmetric Hamiltonian $H$ the time derivative of a probability of the form $\langle n|V|m\rangle= \int dV \psi_n^{\rm L}C_m\psi_m^{\rm R}$ given in \cite{R11e} and \cite{R11ga} can be written as an asymptotic surface term. This probability will be preserved in time if the left- and right-eigenfunctions $\psi_n^{\rm L}$ and $\psi_m^{\rm R}$ of $H$   are well-behaved at large distances. When a non-diagonalizable Hamiltonian is constructed as the limit of a diagonalizable one, the behavior of $\psi_n^{\rm L}$ and $\psi_m^{\rm R}$ is modified  in time but not in space. Thus the asymptotic surface term continues to vanish, and $ \int dV \psi_n^{\rm L}C_m\psi_m^{\rm R}$ continues to be time independent.} 

\section{Constructing the non-diagonalizable case as a limit of a diagonalizable one}
\label{S6}

In the previous section we had shown that for the Jordan-block Hamiltonian given in (\ref{E27}) one could still find a well-defined $V=\sigma_1$ that effects $VHV^{-1}=H^{\dagger}$. However for our example of the Hamiltonian $H=r\cos\theta\sigma_0+ir\sin\theta \sigma_3+s\sigma_1$ given in (\ref{E13}), we found that the $V=G^2$ operator became singular in the Jordan-block limit in which we set $s^2-r^2\sin^2\theta=0$. This is of course to be expected since $G$ brings $H$ to a Hermitian form via $GHG^{-1}=\tilde{H}$, and Jordan-block Hamiltonians cannot be diagonalized. This then raises the question of whether for a Hamiltonian such as $H=r\cos\theta\sigma_0+ir\sin\theta \sigma_3+s\sigma_1$ we could find some other $V$ that effects $VHV^{-1}=H^{\dagger}$ when $s^2-r^2\sin^2\theta \neq 0$, with this latter $V$ then being well-defined in the $s^2-r^2\sin^2\theta\rightarrow 0$ limit, a limit in which it would continue to effect $VHV^{-1}=H^{\dagger}$. As we now show, one can indeed construct such a $V$, with there thus being two classes of operators that effect $VHV^{-1}=H^{\dagger}$ for Hamiltonians whose energy eigenspectra are real and complete, one class being related to the transformation that brings a Hamiltonian to a Hermitian form and the other being related to the Jordan-block limit of a Hamiltonian should it have one.

Immediate inspection of the  Hamiltonian $H=r\cos\theta\sigma_0+ir\sin\theta \sigma_3+s\sigma_1$ shows that this latter class of $V$ operators is not empty, since no matter what the value of $s^2-r^2\sin^2\theta$, the operator $\sigma_1$ effects 
\begin{equation}
\sigma_1H\sigma_1=\sigma_1(r\cos\theta\sigma_0+ir\sin\theta \sigma_3+s\sigma_1)\sigma_1=
r\cos\theta\sigma_0-ir\sin\theta \sigma_3+s\sigma_1=H^{\dagger},
\label{E29}
\end{equation}
doing so not only in both the $s^2-r^2\sin^2\theta>0$ and $s^2-r^2\sin^2\theta<0$ phases but even doing so in the $s^2-r^2\sin^2\theta= 0$ Jordan-block limit as well. To understand why we have found two classes of $V$ operator, we note that when we originally obtained (\ref{E8b}), viz. $B^{\dagger}BHB^{-1}(B^{\dagger})^{-1}=H^{\dagger}$, and then identified $V=B^{\dagger}B$, we could have replaced (\ref{E8b}) by
\begin{equation}
B^{\dagger}BCHCB^{-1}(B^{\dagger})^{-1}=H^{\dagger},
\label{E30}
\end{equation}
since the $C$ operator introduced above obeys $[C,H]=0$ and $C^2=I$. Thus instead of identifying $V=B^{\dagger}B$ we could equally well have identified a new $V^{\prime}$ according to $V^{\prime}=B^{\dagger}BC$, so that in terms of the original $V$  the new $V^{\prime}$ is given  by $V^{\prime}=VC$. However, since for the particular $H$ given in (\ref{E13})  the original $V$ obeyed $V={PC}$, $C=V^{-1}P=PV$ ($P^2$ being equal to one), we thus have 
\begin{equation}
V^{\prime}=VC=P,
\label{E31}
\end{equation}
and thus 
\begin{equation}
PHP=H^{\dagger}.
\label{E32}
\end{equation}
As we see, both $P$ and $PC$ are able to serve as the operator needed to effect $VHV^{-1}=H^{\dagger}$. In the Jordan-block limit $C$, and thus $PC$, is singular while  $P$ is not.  The singular operator ${PC}=V$ is thus associated with the diagonalization of $H$, while the non-singular $P$ can continue to effect $PHP=H^{\dagger}$ even in the Jordan-block  limit. The discussion here thus highlights the utility of the $C$ operator. Starting with a $V^{\prime}=P$ that effects $PHP=H^{\dagger}$, we construct $V=G^2=V^{\prime}C={PC}$, with the operator $G$ then enabling us to bring $H$ to a Hermitian form $GHG^{-1}=\tilde{H}$.

\section{Additional comments}
\label{S7}
\subsection{$PT$ symmetry and path integration}

In this paper we have shown how $PT$ invariance of a Hamiltonian is intimately connected with unitary time evolution. We have identified three distinct ways in which a $PT$ symmetry of a Hamiltonian  can be implemented, viz. the energy eigenspectrum of the Hamiltonian is real and complete, the energy eigenspectrum contains complex conjugate pairs but is still complete, the energy eigenspectrum is real but incomplete. In each case we have shown that  time evolution is unitary. In all of these cases the crucial common ingredient is not the reality of the energy eigenvalues but the reality of the secular equation that determines them. Since any non-real secular equation would have to possess at least one complex eigenvalue that is not part of a complex conjugate pair, the analysis of (\ref{E3}) and (\ref{E4}) shows that  time evolution could not then be unitary. Reality of the secular equation is thus the necessary and sufficient condition for unitary time evolution.

Given the centrality of the reality of the secular equation itself rather than the reality of its solutions, one not only recognizes $PT$ symmetry as having primacy over Hermiticity, one has to then ask how it is that Hermiticity actually comes into physics in the first place. As a requirement on quantum operators Hermiticity has quite a few shortcomings. Firstly, the primary reason for imposing it is that it leads to real eigenvalues. However, as we have seen, non-Hermitian Hamiltonians can just as easily have real eigenvalues as Hermitian ones, with Hermiticity only being sufficient for reality but not necessary. Secondly, as stressed in \cite{R1}, unlike $PT$ symmetry, Hermiticity is a mathematical rather than a physical requirement. Thirdly, $PT$ symmetry can be imposed at the level of the unconstrained Lagrangian, and when imposed it holds on every path stationary or non-stationary. However, Hermiticity is only imposed on the stationary Hamiltonian, with the Hilbert space only being constructed in the stationary solution. Since the non-stationary paths play a role in path integral quantization, to appreciate the utility of $PT$ symmetry it is instructive to discuss the role that path integrals play in quantum mechanics.

To understand the issues involved we quickly review path integration as it appears in conventional Hermitian quantum mechanics. In Hermitian quantum mechanics one introduces a retarded propagator $G(x^{\prime},t^{\prime};x,t)$ that propagates a wave function $\psi(x,t)$ at time $t$ to a wave function $\psi(x^{\prime},t^{\prime})$ at a later time $t^{\prime}$ according to $\psi(x^{\prime},t^{\prime})=i\int dx G(x^{\prime},t^{\prime};x,t)\psi(x,t)$. If $\psi(x,t)$ obeys the Schr\"odinger equation $i\partial_t\psi(x,t)=H(x,-i\partial_x,t)\psi(x,t)$, then $G(x^{\prime},t^{\prime};x,t)$  obeys the propagator equation 
\begin{equation}
[i\partial_{t^{\prime}}-H(x^{\prime},-i\partial_{x^{\prime}},t^{\prime})]G(x^{\prime},t^{\prime};x,t)=\delta(x-x^{\prime})\delta(t-t^{\prime}).
\label{E32a}
\end{equation} 
In Hermitian quantum mechanics there are various equivalent representations for this  propagator. When the Hamiltonian is Hermitian, one can introduce a complete set of energy eigenfunctions $u_i(x)$ with energy $E_i$ that obey $\int dx u^*_i(x)u_j(x)=\delta_{i,j}$, $\sum _iu^*_i(x^{\prime})u_i(x)=\delta(x-x^{\prime})$. In terms of these eigenfunctions one can represent the propagator as $G(x^{\prime},t^{\prime};x,t)=-i\theta(t^{\prime}-t)\sum_iu^*_i(x^{\prime})u_i(x)e^{-iE_i(t^{\prime}-t)}$, and one can verify directly that it obeys the propagator equation (\ref{E32a}). A second way to represent the propagator is as the matrix element $-i\theta(t^{\prime}-t)\langle x,t| x^{\prime}t^{\prime}\rangle=-i\theta(t^{\prime}-t)\langle x|e^{-iH(t^{\prime}-t)}|x^{\prime}\rangle$, since the insertion of $\sum_i|i\rangle\langle i|=I$ as summed over a complete basis of energy eigenstates and the identifications $u_i(x)=\langle x|i\rangle$, $u^*_i(x^{\prime})=\langle i|x^{\prime}\rangle$  leads straight back to $G(x^{\prime},t^{\prime};x,t)=-i\theta(t^{\prime}-t)\sum_iu^*_i(x^{\prime})u_i(x)e^{-iE_i(t^{\prime}-t)}$. 

A third way to represent the propagator is given via the Feynman path integral. Here one breaks up the time interval $t^{\prime}-t$ into infinitesimal time slices and introduces complete sets of position and momentum eigenstates at each time slice. Thus for the Hamiltonian $H=\hat{p}^2/2m +V(\hat{x})$ where $\hat{p}$ and $\hat{x}$ are the momentum and position operators, we set $t^{\prime}=t+\epsilon$, and to lowest order in $\epsilon$ we straightforwardly obtain 
\begin{eqnarray}
G(x^{\prime},t+\epsilon;x,t)&=&-i\theta(\epsilon)\langle x|\left[1-i\epsilon\frac{\hat{p}^2}{2m} -i\epsilon V(\hat{x})\right]|x^{\prime}\rangle
\nonumber\\
&=&-i\theta(\epsilon)\left[\delta(x^{\prime}-x)[1 -i\epsilon \bar{V}(x,x^{\prime})]
-i\epsilon\int \frac{dp}{2\pi}\frac{p^2}{2m}e^{ip(x^{\prime}-x)}\right]
\nonumber\\
&=&-i\theta(\epsilon)\int \frac{dp}{2\pi}e^{ip(x^{\prime}-x)}\left[1-i\epsilon\frac{p^2}{2m} -i\epsilon \bar{V}(x^{\prime}x)\right]
\nonumber\\
&=&-i\theta(\epsilon)\int \frac{dp}{2\pi}\exp\left[ip(x^{\prime}-x)-i\epsilon\frac{p^2}{2m}-i\epsilon\bar{V}(x^{\prime}x)\right]
\nonumber\\
&=&-i\theta(\epsilon)\left(\frac{m}{2\pi i \epsilon}\right)^{1/2}\exp\left[\frac{im(x^{\prime}-x)^2}{2\epsilon}-i\epsilon\bar{V}(x^{\prime}x)\right].
\label{E32b}
\end{eqnarray}
In (\ref{E32b}) we have represented the matrix element $\langle x|V(\hat{x})|x^{\prime}\rangle$ as some average value $\bar{V}(x,x^{\prime})$ of the potential in the $(x,x^{\prime})$ interval. Now while it is tempting to identify $m(x^{\prime}-x)^2/2\epsilon^2-\bar{V}(x^{\prime},x)$ as the classical Lagrangian $L_{\rm CL}$ (and thus identify $\epsilon L_{\rm CL}$ as the classical action $S_{\rm CL}$), one cannot initially make such an identification since $m(x^{\prime}-x)^2/2\epsilon^2-\bar{V}(x^{\prime},x)$ would only be the classical Lagrangian appropriate to the infinitesimal time interval $\epsilon$ if $x^{\prime}$ were infinitesimally close to $x$.  However, the propagator at time $t^{\prime\prime}=t+2\epsilon$ is given by $G(x^{\prime\prime},t^{\prime \prime};x,t) =i\int dx^{\prime}G(x^{\prime\prime},t^{\prime \prime};x^{\prime},t+\epsilon) G(x^{\prime},t+\epsilon;x,t)$, to thus involve an integration over all possible values of $x^{\prime}$ and not just those infinitesimally close to $x$. Nonetheless, as noted for instance in \cite{R19}, the presence of the oscillating $e^{im(x^{\prime}-x)^2/2\epsilon}$ factor in (\ref{E32b}) then suppresses all values of $x^{\prime}$ that are not  infinitesimally close to $x$. Thus, as long as we perform the subsequent $x^{\prime}$ integration we can replace $\bar{V}(x^{\prime},x)$ by $V(x)$ and set 
\begin{equation}
G(x^{\prime},t+\epsilon;x,t)=-i\theta(\epsilon)\left(\frac{m}{2\pi i \epsilon}\right)^{1/2}e^{iS_{\rm CL}(x^{\prime},x)}.
\label{E32c}
\end{equation}
Iterating $G(x^{\prime\prime},t^{\prime \prime};x,t) =i\int dx^{\prime}G(x^{\prime\prime},t^{\prime \prime};x^{\prime},t+\epsilon) G(x^{\prime},t+\epsilon;x,t)$ to a full $(t,t^{\prime}$) time interval then leads to the Feynman path integral formula in which one can represent $G(x^{\prime},t^{\prime};x,t)$ as $-i\theta(t^{\prime}-t)\int D[x]e^{iS_{\rm CL}[x]}$ as summed over all classical paths that originate at $x$ at time $t$ and reach $x^{\prime}$ at time $t^{\prime}$, with $S_{\rm CL}[x]$ being the value that the classical action takes in each such path. 

Now suppose we allow for the possibility that the Hamiltonian is not Hermitian. Now in this case  we can still represent the propagator as the matrix element $-i\theta(t^{\prime}-t)\langle x|e^{-iH(t^{\prime}-t)}|x^{\prime}\rangle$, since the Hamiltonian is still the generator of time translations. And again we can still insert complete sets of energy eigenstates or complete sets of momentum eigenstates. For energy eigenstates we must set $\sum_i|R_i\rangle \langle R_i|V=\sum_i|R_i\rangle \langle L_i|=I$, and on making the identifications $\psi^{{\rm R}}_i(x, t)=\langle x|R_{i}(t)\rangle$, $\psi^{{\rm L}}_i(x, t)=\langle L_{i}(t)|C|x\rangle=C_i\langle L_{i}(t)|x\rangle$ introduced above, we obtain 
$G(x^{\prime},t^{\prime};x,t)=-i\theta(t^{\prime}-t)\sum_i\psi_i^{\rm R}(x)C_i\psi^{\rm L}_i(x^{\prime})e^{-iE_i(t^{\prime}-t)}$. For momentum eigenstates, the momentum eigenstates will still be complete, but with $p$ and $x$ possibly no longer being Hermitian,  the $[\hat{x},\hat{p}]=i$ commutator would instead be represented \cite{R11c} by $[e^{i\theta}x,-ie^{-i\theta}\partial_x]=i$ where $\theta$ is an appropriate phase, and the position operator eigenvalues $x$ and $x^{\prime}$ that label the states $|x\rangle$ and $|x^{\prime}\rangle$ would no longer be real.\footnote{Given only that $[\hat{x},\hat{p}]=i$, the operator $\hat{S}(a)=e^{-ia\hat{p}}$ with constant $a$ obeys $[\hat{x},\hat{S}(a)]=a\hat{S}(a)$. Thus if $|x\rangle$ is an eigenstate of $\hat{x}$ with eigenvalue $x$, then $\hat{S}(a)|x\rangle$ is an eigenstate of $\hat{x}$ with eigenvalue $x+a$. Since this analysis holds for arbitrary $a$, the eigenspectrum of $\hat{x}$ is continuous. Moreover, since it holds independent of whether $a$ is real or complex, for non-zero $\theta$ the eigenspectrum of the position operator will be complete on an appropriate contour in the complex coordinate plane. Similarly, since the operator $\hat{T}(b)=e^{ib\hat{x}}$ obeys $[\hat{p},\hat{T}(b)]=b\hat{T}(b)$, the eigenspectrum of the momentum operator will equally be complete on an appropriate contour in the complex momentum plane.} In this case (\ref{E32b}), (\ref{E32c}) and the Feynman path integral formula will all still hold, but now as evaluated on the relevant complex coordinate space paths. As we see, using the momentum eigenstate approach we can derive the path integral formula without needing to know what the closure relation for energy eigenstates might be, and thus without needing to make any reference to the operator $V$ at all.

While the above approach yields the path integral formula as output, i.e. start with a quantum theory in a Hilbert space and evaluate $-i\theta(t^{\prime}-t)\langle x|e^{-iH(t^{\prime}-t)}|x^{\prime}\rangle$, one can instead start with the path integral as input and define quantization as given by the path integral. In this latter  approach one can define the quantum theory once and for all by defining the propagator as $G(x^{\prime},t^{\prime};x,t)=-i\theta(t^{\prime}-t)\int D[x]e^{iS_{\rm CL}[x,x^{\prime}]}$. The utility of this latter approach is that one does not need to preassign a Hilbert space. Rather, it is only after one has constructed  $G(x^{\prime},t^{\prime};x,t)$ via a path integral that one then seeks to identify a Hilbert space and appropriate Hamiltonian in which $G(x^{\prime},t^{\prime};x,t)$ can be represented as a matrix element of the form $i\theta(t^{\prime}-t)\langle x|e^{-iH(t^{\prime}-t)}|x^{\prime}\rangle$. In this latter case the associated Hilbert space norm is output rather than input, and for an appropriate $-i\theta(t^{\prime}-t)\int D[x]e^{iS_{\rm CL}[x,x^{\prime}]}$ the associated Hilbert space could be the one associated with the $PT$ theory norm rather than the one associated with the standard Dirac norm.

To be more specific, if we quantize a theory via path integration, i.e. if we define a quantum theory via a path integral, then since the path integration is only over classical paths, there is no reference in it at all to Hermiticity or even to a Hilbert space for that matter. Since path integration quantization is a pure c-number approach, from it one constructs not quantum operators themselves but only their correlators or Green's functions, i.e. one constructs  c-number matrix elements of quantum operators as evaluated in some appropriate Hilbert space basis states. Hermiticity can thus enter physics if one is able to construct these same matrix elements using Hermitian operators acting on a standard Dirac-norm Hilbert space. If one cannot construct matrix elements this way then one is in a non-Hermitian quantum theory. Thus one has to ask what it is in a path integral that would determine whether or not the associated quantum theory is a standard Dirac-norm Hermitian theory.

While we have yet to provide a complete answer to this question, the answer  would appear to lie in the nature of the domain of the path integral measure that is needed for the path integral to actually exist. From the cases that have so far been studied, a pattern has emerged. Specifically, for standard Hermitian theories the domain is over real paths, while for non-Hermitian ones the needed domain is over complex ones. When the path integral exists for a real basis of classical paths, we are able to identify classical operators as the real eigenvalues of Hermitian quantum operators, with path integration then being equivalent to standard canonical quantization. However, in $PT$ theories the needed domains are in the complex plane in so-called Stokes wedges \cite{R1} where the asymptotic behavior of the theory is under control. For theories where the Hamiltonian is neither Hermitian nor $PT$ symmetric  (i.e. theories in which there are decaying modes but no complex conjugate growing ones), experience with pair production in an external electromagnetic field in the presence of instantons \cite{R14} shows that the semi-classical approximation to the path integral also requires complex paths.\footnote{The author is indebted to Dr. G.~V.~Dunne for informing him of this reference.}  It thus appears that Hermiticity of operators is associated with path integral measures that are real (measures that could of course be $PT$ symmetric as well), while theories whose path integral measures only exist in the complex plane have a much more general structure.

If one can associate a real path integral measure with Hermitian theories because Hermitian operators have real eigenvalues, and if one needs to use a complex path integral measure in the general non-Hermitian case, one has to ask what would be the indicator that one is in a non-Hermitian case that is $PT$ symmetric. The answer to this would appear to be that in the $PT$ case the Wick-rotated Euclidean path integral would (if the continuation exists of course) be real. Specifically, since the secular equation for the eigenvalues is real in the $PT$ case, the eigenvalues are either real or appear in complex conjugate pairs, and when they do appear in complex conjugate pairs they appear with complex conjugate wave functions \cite{R1}. Consequently, for real or complex conjugate pairs the Euclidean path integral, which following \cite{R11e,R11ga} and our analysis above behaves as $\sum_n \psi_n^{\rm R}(x)C_n\psi_n^{\rm L}(x^{\prime})e^{-E_n\tau}$ at Euclidean time $\tau$, is real. Thus a real measure and a real Euclidean path integral is associated with a Hermitian theory, while a complex path integral measure but a real Euclidean path integral is associated with a non-Hermitian but $PT$-symmetric  theory.

\subsection{Symplectic symmetry in classical mechanics}

In the above we had noted that the very existence of a path integral is determined by the domain of paths in a complex classical space for which the  integration over the paths yields a finite answer. Also we had noted that there was an invariance of the quantum-mechanical commutation relation $[x,p]=i$ to complex transformations of the form $x \rightarrow e^{i\theta}x$, $p \rightarrow e^{-i\theta}p$. In this section we relate these two issues by showing that a complex invariance structure can also be found for Poisson brackets in classical mechanics. 

To be specific, we consider a classical system with $n$ coordinates $q_i$, $n$ momenta $p_i$, and generic Poisson bracket
\begin{equation}
\{u,v\}={\displaystyle \sum}_{i=1}^{i=n}\left(\frac{\partial u}{\partial q_i}\frac{\partial v}{\partial p_i}-\frac{\partial u}{\partial p_i}\frac{\partial v}{\partial q_i}\right).
\label{A17}
\end{equation}
If we introduce a $2n$-dimensional vector $\eta$ and a $2n$-dimensional matrix $J$ defined as
\begin{eqnarray}
\eta=\pmatrix{q_i\cr p_i},~~~~J=\pmatrix{0&I\cr -I&0},
\label{A18}
\end{eqnarray}
where $I$ is an $n$-dimensional unit matrix, we can compactly write the generic Poisson bracket as 
\begin{equation}
\{u,v\}=\widetilde{\frac{\partial u}{\partial \eta}}J\frac{\partial v}{\partial \eta},
\label{A19}
\end{equation}
where the tilde symbol denotes transpose. If we now make a phase space transformation to a new $2n$-dimensional vector $\xi$ according to 
\begin{equation}
M_{ij}=\frac{\partial \xi_i}{\partial \eta_j},\qquad \frac{\partial v}{\partial \eta}=\tilde{M}\frac{\partial v}{\partial \xi},\qquad \widetilde{\frac{\partial u}{\partial \eta}}=\widetilde{\frac{\partial u}{\partial \xi}}M,
\label{A20}
\end{equation}
then in these new coordinates and momenta the Poisson bracket takes the form
\begin{equation}
\{u,v\}=\widetilde{\frac{\partial u}{\partial \xi}}MJ\tilde{M}\frac{\partial v}{\partial \xi}.
\label{A21}
\end{equation}
The Poisson bracket will thus be left invariant for any  $M$ that obeys the symplectic symmetry relation
\begin{equation}
MJ\tilde{M}=J.
\label{A22}
\end{equation}

If we introduce generators $G$ defined according to $M=e^{i\omega G}$, such generators will obey 
\begin{equation}
e^{i\omega G}Je^{i\omega \tilde{G}}=J,~~~~GJ+J\tilde{G}=0.
\label{A23}
\end{equation}
Since the matrix $J$ obeys $J^{-1}=\tilde{J}=-J$, the generators will obey
\begin{equation}
\tilde{G}=-J^{-1}GJ=JGJ.
\label{A24}
\end{equation}
Solutions to $\tilde{G}=JGJ$ can be broken into two classes, symmetric generators that  anticommute with $J$, viz. those that obey
\begin{equation}
\tilde{G}=G,\qquad GJ+JG=0,
\label{A25}
\end{equation}
and symmetric generators that commute with $J$, viz. those that obey

\begin{equation}
\tilde{G}=-G,\qquad GJ-JG=0.
\label{A26}
\end{equation}
In $N=2n$ dimensions there are $N(N-1)/2$ symmetric generators and $N$ antisymmetric generators, for a total of $N(N+1)/2$ generators. These $N(N+1)/2$ generators close on the Lie algebra $Sp(N)$, the symplectic group in $2n$ dimensions. With the generic Lie algebra commutation relations being of the form $[G_i,G_j]=i\sum_k f_{ijk}G_k$ with real structure coefficients $f_{ijk}$, one can find representations of the symplectic algebra in which all the generators are pure imaginary. Thus if, as is standard in classical mechanics, one takes all angles $\omega$ to be real, canonical transformations effected by $e^{i\omega G}$ will transform a real $\eta$ into a real $\xi$.

However, since the algebra of the generators makes no reference to angles, invariance of the classical Poisson brackets under canonical transformations will persist even if the $\omega$ are taken to be complex. The Poisson brackets of classical mechanics thus possesses a broader class of invariances than those associated with real canonical transformations alone since one can transform a real $\eta$ into a complex $\xi$ and still preserve the Poisson bracket algebra. Consequently, with both the classical Poisson brackets and the quantum commutators admitting of complex canonical transformations, for every such complex transformation we are able to construct a canonical quantization with an associated correspondence principle. Namely, for each canonically transformed Poisson bracket we associate a canonically transformed quantum commutator, with each associated set of classical coordinates being the eigenvalues of the associated transformed quantum operators. 

Thus, with classical mechanics possessing complex symplectic invariances and not just real ones, we can generalize canonical quantization and the correspondence principle to complex coordinates. Thus instead of thinking of classical mechanics as being based on real numbers, we should think of it as being based on c-numbers, viz. numbers that commute with each other but are not necessarily real. Moreover, if we extend classical physics to a Grassmann space in which there are Grassmann numbers that anticommute with each other, such Grassmann numbers would also be c-numbers, and would become q-numbers when vanishing anticommutators are replaced by non-vanishing ones. In such Grassmann spaces there would be an intrinsic dependence on the square root of minus one, with complex numbers thus being natural in a purely classical world.\footnote{A familiar example of this would be the Majorana mass $\sum \tilde{\psi}_{\alpha}C_{\alpha\beta}\psi_{\beta}$ where the matrix $C_{\alpha\beta}$ is the Dirac transposition matrix that effects $C^{-1}\gamma_{\mu}C=-\tilde{\gamma}_{\mu}$. With $C_{\alpha\beta}$ being antisymmetric,  in the Majorana basis for the Dirac gamma matrices where the components of the Majorana spinor obey $\psi^{\dagger}=\psi$, the Majorana mass takes the form (see e.g. \cite{R11ff,R11fg}) $\tilde{\psi}C\psi=-i\psi_1\psi_4+i\psi_2\psi_3-i\psi_3\psi_2+i\psi_4\psi_1$. Thus $\tilde{\psi}C\psi$  is only non-zero if the components of the Majorana spinor obey the anticommuting $\psi_{\alpha}\psi_{\beta}+\psi_{\beta}\psi_{\alpha}=0$, and is only Hermitian if the square root of minus one factor is present. A quantization of the Grassmann theory then replaces vanishing anticommutators by non-vanishing ones, with the quantum-mechanical Dirac propagator being derivable \cite{R11fg} via  a path integration over Grassmann paths.} Thus the general definition of the  classical limit of quantum mechanics is to replace q-numbers by c-numbers and not by real ones.

Now while classical mechanics contains this broad class of complex symplectic transformations, they ordinarily play no role in physics since they contain no additional information that is not already contained in the real symplectic transformations alone. However, this would not be the case if we were to encounter some form of discontinuity when we continue into the complex coordinate plane. In path integral quantization these discontinuities would occur if the path integral only existed for  a domain of classical paths that were not real. In a canonical quantization these discontinuities would occur if quantum-mechanical wave functions are convergent in the domain associated with some Stokes wedges and divergent in some other Stokes wedges. If the domain of convergence includes the real coordinate axis we are in conventional Hermitian quantum mechanics, and we can take the classical limit to be based on real numbers. However, if wave functions are only convergent in Stokes wedges that do not include the real axis, we are in a non-Hermitian realization of the theory. Now amongst such general non-Hermitian realizations there will be some that are also $PT$ realizations. We can therefore anticipate that the ones that are $PT$ realizations are those in which the classical domain for which the path integral exists and the quantum-mechanical domain for which wave functions exist are either the same or related in some way.

To conclude this section we note that for the simple case of a 4-dimensional phase space (viz. $n=2$) the 4-dimensional transposition matrix $C$ that  is involved in charge conjugation of Dirac spinors and effects $C^{-1}\gamma_{\mu}C=-\tilde{\gamma}_{\mu}$ can also play a role in symplectic transformations. In order to be able to use a $C$ that is, like $J$, antisymmetric, orthogonal, and composed of real elements alone, we take $C$ to be given by its representation in the Weyl basis of the Dirac gamma matrices, viz.
\begin{eqnarray}
C=\pmatrix{-i\sigma_2&0\cr 0&i\sigma_2}.
\label{A27}
\end{eqnarray}
Then, if we now take $\eta$ to be of the form   
\begin{eqnarray}
\eta=\pmatrix{p_1\cr q_1\cr q_2 \cr p_2\cr},
\label{A28}
\end{eqnarray}
we can write the Poisson bracket as
\begin{equation}
\{u,v\}=\widetilde{\frac{\partial u}{\partial \eta}}C\frac{\partial v}{\partial \eta},
\label{A29}
\end{equation}
with the symplectic condition then being given by
\begin{equation}
MC\tilde{M}=C.
\label{A30}
\end{equation}
Moreover, since (\ref{A30}) would not be affected if we were to replace $C$ by $iC$, we would then have an operator $iC$ whose square is one, to thus be reminiscent of the $PT$ theory $C$ operator described earlier. In such a case it would be $i\{u,v\}$ that would be defined as the Poisson bracket, and under a canonical quantization the quantum commutator $[\hat{u},\hat{v}]$ would be identified with  $\hbar$ times it.

\subsection{$PT$ symmetry and the Pais-Uhlenbeck fourth-order oscillator}

In Sec. (7.1) we had discussed the possibility that for $PT$ symmetric theories the path integral measure would be complex and its Euclidean extension would be real. Here we briefly describe how this possibility is realized in the completely solvable fourth-order Pais-Uhlenbeck oscillator theory \cite{R15} studied in \cite{R11c,R11d,R11e,R16}. For this theory the action is given by 
\begin{equation}
I_{\rm PU}=\frac{\gamma}{2}\int dt\left[{\ddot
z}^2-\left(\omega_1^2+\omega_2^2
\right){\dot z}^2+\omega_1^2\omega_2^2z^2\right],
\label{E33}
\end{equation}
where there are two frequencies $\omega_1$ and $\omega_2$ and an overall coefficient $\gamma$ that is a positive constant. Since the action depends on three dynamical degrees of freedom (position, velocity, and acceleration), the theory is constrained since the velocity has to serve as the canonical conjugate of both the position and the acceleration. To rectify this one replaces $\dot{z}$ by a new variable $x$, and via the method of Dirac constraints constructs \cite{R17,R18} the Hamiltonian  
\begin{equation}
H_{\rm PU}=\frac{p_x^2}{2\gamma}+p_zx+\frac{\gamma}{2}\left(\omega_1^2+
\omega_2^2\right)x^2-\frac{\gamma}{2}\omega_1^2\omega_2^2z^2,
\label{E34}
\end{equation}
as based on two canonical pairs of variables $(z,p_z)$ and $(x,p_x)$. If one replaces $\omega_1^2$ and $\omega_2^2$ by $\omega_1^2-i\epsilon$ and $\omega_2^2-i\epsilon$, one finds \cite{R11e} that because of the $z^2$ term, the $\int[dz]e^{iI_{\rm PU}}$ path integral does not exist in paths in which $z$ is real. The path integral does however exist in the complex $z$ plane in Stokes wedges that contain the imaginary $z$ axis. The Pais-Uhlenbeck theory thus has to be reinterpreted as a $PT$ theory with a non-Hermitian, but $PT$-symmetric, Hamiltonian $H_{\rm PU}$, and in consequence the Pais-Uhlenbeck theory is unitary and free of any negative Dirac-norm ghost states \cite{R11c}. The eigenvalues of $H_{\rm PU}$ are real, with its potentially unbounded from below $-\gamma\omega_1^2\omega_2^2z^2/2$ term becoming bounded when $z$ is anti-Hermitian. (The $\gamma(\omega_1^2+
\omega_2^2)x^2/2$ term will also be bounded from below if, unlike $z$, $x$ is Hermitian \cite{R11c}.) Since the energies are real and bounded from below, the deep Euclidean path integral is both real and finite (no $e^{-E_n\tau}$ terms with $E_n<0$), just as found in \cite{R16}.

The  Pais-Uhlenbeck fourth-order oscillator can be generalized to quantum field theory where it is then based on the generic action 
\begin{equation}
I_{\rm S}=-\frac{1}{2}\int d^4x\left[\partial_{\mu}\partial_{\nu}\phi\partial^{\mu}
\partial^{\nu}\phi+(M_1^2+M_2^2)\partial_{\mu}\phi\partial^{\mu}\phi+M_1^2M^2_2\phi^2\right],
\label{M23}
\end{equation}
with a generic scalar field $\phi(x)$. (The $I_{\rm PU}$ action is the limit of $I_{\rm S}$ in configurations with a fixed momentum $\bar{k}$ in which $\omega_1^2=\bar{k}^2+M_1^2$,  $\omega_2^2=\bar{k}^2+M_2^2$.) Given this action one obtains an equation of motion
\begin{equation}
(-\partial_t^2+\bar{\nabla}^2-M_1^2)(-\partial_t^2+\bar{\nabla}^2-M_2^2)\phi(x)=0,
\label{M24}
\end{equation}
a propagator of the form
\begin{equation}
D(k,M_1,M_2)=\frac{1}{(M_2^2-M_1^2)}\left(\frac{1}{k^2+M_1^2}-\frac{1}{k^2+M_2^2}\right),
\label{M25}
\end{equation}
a Hamiltonian of the form $H=\int d^3x T_{00}(M_1,M_2)$
where
\begin{equation}
T_{00}(M_1,M_2)=\pi_{0}\dot{\phi}+\frac{1}{2}\left[\pi_{00}^2+(M_1^2
+M_2^2)(\dot{\phi}^2-\partial_{i}\phi\partial^{i}\phi)-M_1^2M_2^2\phi^2
-\pi_{ij}\pi^{ij}\right],
\label{M26}
\end{equation}
and canonical conjugates that obey
\begin{eqnarray}
&&\pi^{\mu}=\frac{\partial{L}}{\partial \phi_{,\mu}}-\partial_{\lambda
}\left(\frac{\partial {L}}{\partial\phi_{,\mu,\lambda}}\right)=-(M_1^2+M_2^2)\partial^{\mu}\phi+\partial_{\lambda}\partial^{\mu}\partial^{\lambda}\phi,~~~\pi^{\mu\lambda}=\frac{\partial {L}}{\partial \phi_{,\mu,\lambda}}=-\partial^{\mu}\partial^{\lambda}\phi,
\nonumber\\
&&[\phi(\bar{x},t),\pi_0(\bar{x}^{\prime},t)]=i\hbar\delta^3(\bar{x}-\bar{x}^{\prime}),~~~[\partial_0\phi(\bar{x},t),\pi^{0}_{\phantom{0}0}(\bar{x}^{\prime},t)]=i\hbar\delta^3(\bar{x}-\bar{x}^{\prime}).
\label{M27}
\end{eqnarray}
Because of the $-(M_1^2+M_2^2)\partial_{i}\phi\partial^{i}\phi-M_1^2M_2^2\phi^2$ term the Hamiltonian would be unbounded from below if the scalar field is Hermitian. However, as shown in \cite{R11e}, the theory has a realization in which the scalar field is anti-Hermitian, and in this realization the energy eigenspectrum is bounded from below. In this realization one has to use the $\langle R_j(t)|V|R_i(t)\rangle$ norm, with this norm accounting for the relative minus sign in $D(k,M_1,M_2)$, with the theory consequently being unitary and ghost free. As we see, $PT$-symmetry not only takes care of the unitarity of time evolution, it also takes care of the negative Dirac norm ghost states by using a norm other than the Dirac one. 

\subsection{$PT$ symmetry and stability}

The quantum field-theoretic generalization of the Pais-Uhlenbeck fourth-order oscillator is also of interest for another reason. As written, the fourth-order equation of motion given in (\ref{M24}) admits of both positive and negative frequency solutions, and thus the fourth-order theory could potentially possess negative energy instabilities. Now we recall that an analogous situation is encountered in the second-order case where the second-order Klein-Gordon wave equation also possesses both positive and negative frequency solutions. However despite this,  the Klein-Gordon theory is stable because its Hamiltonian (a conventionally Hermitian Hamiltonian) is positive definite, so that the allowed energy eigenstates are associated with the  positive frequency modes alone.  Now, in and of itself, this same stabilization technique does not initially work for the fourth-order case, since not only does the wave equation have negative frequency solutions, if all the fields of the fourth-order theory are Hermitian, the Hamiltonian associated with (\ref{M26}) would have negative energy eigenstates and the theory would be unstable. However, if we take the scalar field, and hence its $\pi_0$ conjugate as well,  to be anti-Hermitian rather than Hermitian, we can then reinterpret the Hamiltonian as a positive definite $PT$-symmetric Hamiltonian,\footnote{In configurations with zero linear momentum the $-(1/2)M_1^2M_2^2\phi^2$ term in (\ref{M26})  acts analogously to the $-(\gamma/2)\omega_1^2\omega_2^2z^2$ term in (\ref{E34}).} and can do so without jeopardizing unitarity, with the theory then having no states of negative energy and no states of negative norm. We thus recognize  $PT$ symmetry as the mechanism that generalizes   the second-order Klein-Gordon Hamiltonian stabilization  procedure to higher-derivative theories.

\appendix
\setcounter{equation}{0}
\def\theequation{A\arabic{equation}}
\section{Distinguishing $\hat{H}$ and $\tilde{H}$ and the lack of uniqueness of $V$}
\label{S8}

To illustrate the distinction between the Hamiltonians $\hat{H}$ and $\tilde{H}$ introduced in Sec. (3) we take as $\hat{H}$ the operator 
\begin{equation}
\hat{H}=a_0\sigma_0+a_1\sigma_1+a_2\sigma_2+a_3\sigma_3,
\label{A1}
\end{equation}
with $\hat{H}$ being Hermitian if the otherwise arbitrary coefficients $a_0,a_1,a_2,a_3$ are all real. For illustrative purposes we introduce a $B$ operator of the form 
\begin{equation}
B=\frac{1}{i\surd{3}}(\sigma_0+2\sigma_1),~~~B^{-1}=\frac{1}{i\surd{3}}(\sigma_0-2\sigma_1).
\label{A2}
\end{equation}
We have selected this particular $B$ since for our purposes here we need only consider a $B$ that is general enough to be neither Hermitian nor positive, with the eigenvalues of this particular $B$ being $-3i/\surd{3}$ and $i/\surd{3}$. Given this $B$, we construct $H=B^{-1}\hat{H}B$, to obtain the non-Hermitian
\begin{equation}
H=a_0\sigma_0+a_1\sigma_1-\frac{\sigma_2}{3}(5a_2+4ia_3)
-\frac{\sigma_3}{3}(5a_3-4ia_2).
\label{A3}
\end{equation}
From $B$ we construct $V=B^{\dagger}B$ and obtain 
\begin{equation}
V=\frac{1}{3}(5+4\sigma_1),~~~V^{-1}=\frac{1}{3}(5-4\sigma_1),
\label{A4}
\end{equation}
and with the eigenvalues of $V$ being $3$ and $1/3$ confirm that $V$ is a positive, Hermitian operator. With this $V$ we  obtain
\begin{equation}
VHV^{-1} =a_0\sigma_0+a_1\sigma_1-\frac{\sigma_2}{3}(5a_2-4ia_3)
-\frac{\sigma_3}{3}(5a_3+4ia_2),
\label{A5}
\end{equation}
and confirm that $VHV^{-1}=H^{\dagger}$.
On setting $V=G^2$ we construct an operator
\begin{equation}
G=\frac{1}{\surd{3}}(2\sigma_0+\sigma_1),~~~G^{-1}=\frac{1}{\surd{3}}(2\sigma_0-\sigma_1)
\label{A6}
\end{equation}
with $G$ being a Hermitian operator with positive eigenvalues  $3/\surd{3}$ and $1/\surd{3}$. With this $G$ we construct  
\begin{equation}
GHG^{-1}=\tilde{H}=a_0\sigma_0+a_1\sigma_1-a_2\sigma_2-a_3\sigma_3,
\label{A7}
\end{equation}
and confirm that $\tilde{H}$ is Hermitian. Finally we evaluate $BG^{-1}=-i\sigma_1$ and confirm that the operator $BG^{-1}$ that effects 
\begin{equation}
BG^{-1}\tilde{H}GB^{-1}=\hat{H}
\label{A8}
\end{equation}
is unitary, just as required.

In the above construction we started with a general Hermitian $\hat{H}$ and applied a similarity transform on it to bring it to the non-Hermitian $H=B^{-1}\hat{H}B$. Now there was nothing special about the $B$ that we chose, and we could just as well have used any $B$. For any choice of Hamiltonian that can be written as  $H=B^{-1}\hat{H}B$ where $\hat{H}$ is Hermitian, there must exist a positive, Hermitian operator $G$ that effects $GHG^{-1}=\tilde{H}$ where $\tilde{H}$ is Hermitian too.

It is also of interest to explicitly construct the $P$, $T$ and $C$ operators for this model. If we first write $H$ in the generic form 
\begin{equation}
H=h_0\sigma_0+\mbox{\boldmath $\sigma$}\cdot {\bf h}
\label{A9}
\end{equation}
with energies $E_{\pm}=h_0\pm ({\bf h}\cdot {\bf h})^{1/2}$, the energies will be real if $h_0$ is real and ${\bf h}\cdot {\bf h}$ is real and non-negative, i.e. if the real and imaginary parts of ${\bf h}={\bf h_{\rm R}}+i{\bf h_{\rm R}}$ obey ${\bf h_{\rm R}}{\bf h_{\rm R}}-{\bf h_{\rm I}}{\bf h_{\rm I}}\geq 0$, ${\bf h_{\rm R}}\cdot{\bf h_{\rm I}}=0$. Following \cite{R6}, we define 
\begin{equation}
P=\frac{\mbox{\boldmath $\sigma$}\cdot {\bf h_{\rm R}}}{({\bf h_{\rm R}}\cdot {\bf h_{\rm R}})^{1/2}},~~~
T=K\frac{\sigma_2\mbox{\boldmath $\sigma$}\cdot {\bf h_{\rm R}}\times {\bf h_{\rm I}}}
{({\bf h_{\rm R}}\times {\bf h_{\rm I}}\cdot {\bf h_{\rm R}}\times {\bf h_{\rm I}})^{1/2}},~~~
C=\frac{\mbox{\boldmath $\sigma$}\cdot {\bf h}}{({\bf h}\cdot {\bf h})^{1/2}},
\label{A10}
\end{equation}
where $K$ effects complex conjugation. Then on recalling that $(\mbox{\boldmath $\sigma$}\cdot{\bf A})(\mbox{\boldmath $\sigma$}\cdot
{\bf B})(\mbox{\boldmath $\sigma$}\cdot{\bf A})=2(\mbox{\boldmath $\sigma$}\cdot
{\bf A})({\bf A}\cdot{\bf B})-(\mbox{\boldmath $\sigma$}\cdot{\bf B})({\bf A}
\cdot{\bf A})$,  and that $\sigma_2 \mbox{\boldmath $\sigma$}=-\mbox{\boldmath
$\sigma$}^*\sigma_2$, we readily check that $P^2=I$,  $T^2=I$, $[P,T]=0$, $[PT,H]=0$, $C^2=I$, $[C,H]=0$, $[C,PT]=0$, just as required.  

For these particular $P$, $T$ and $C$ operators we note that $P$ and $PC$ effect
\begin{equation}
PHP=H^{\dagger},~~~PCHCP=H^{\dagger}.
\label{A11}
\end{equation}
In addition we note that from (\ref{A10}) we obtain
\begin{equation}
PC=\frac{({\bf h_{\rm R}}\cdot {\bf h_{\rm R}}
-\mbox{\boldmath $\sigma$}\cdot {\bf h_{\rm R}}\times {\bf h_{\rm I}})}{({\bf h_{\rm R}}\cdot {\bf h_{\rm R}})^{1/2}({\bf h}\cdot {\bf h})^{1/2}}.
\label{A12}
\end{equation}
With the trace and determinant of $PC$ being given by
\begin{equation}
{\rm Tr}[PC]=\frac{2{\bf h_{\rm R}}\cdot {\bf h_{\rm R}}}{({\bf h_{\rm R}}\cdot {\bf h_{\rm R}})^{1/2}({\bf h}\cdot {\bf h})^{1/2}},~~~
{\rm Det}[PC]=I,
\label{A13}
\end{equation}
we see that $PC$ is a positive, Hermitian operator, just as required.

Specializing now to the particular Hamiltonian given in (\ref{A3}), we identify 
\begin{equation}
{\bf h_{\rm R}}=\left(a_1,-\frac{5a_2}{3},-\frac{5a_3}{3}\right),~~~{\bf h_{\rm I}}=\left(0,-\frac{4a_3}{3},\frac{4a_2}{3}\right),
\label{A14}
\end{equation}
and confirm that ${\bf h_{\rm R}}\cdot {\bf h_{\rm I}}=0$, and that ${\bf h}\cdot {\bf h}=a_1^2+a_2^2+a_3^2$ is both real and non-negative. Since $V$ effects $VHV^{-1}=H^{\dagger}$ we see that $PV$ commutes with $H$. While $PV$ could potentially serve as the $C$-operator, to do so it would need to square to unity. However, on evaluating  $PV$ we obtain 
\begin{equation}
PV=\frac{1}{3({\bf h_{\rm R}}\cdot{\bf h_{\rm R}})^{1/2}}(4h^{\rm R}_1+5\mbox{\boldmath $\sigma$}\cdot {\bf h}),
\label{A15}
\end{equation}
with $(PV)^2$ thus not being equal to one. To rectify this we note that any operator of the form $J=a\sigma_0+b\mbox{\boldmath $\sigma$}\cdot {\bf h}$ will commute with $H=h_0\sigma_0+\mbox{\boldmath $\sigma$}\cdot {\bf h}$. The operator $PVJ$ will thus commute with $H$ as well. On choosing
\begin{eqnarray}
a&=&-\frac{5b{\bf h}\cdot {\bf h}}{4h^{\rm R}_1}=-\frac{5b(a_1^2+a_2^2+a_3^2)}{4a_1},
\nonumber\\
b&=&\frac{12h^{\rm R}_1({\bf h_{\rm R}}\cdot {\bf h_{\rm R}})^{1/2}}{[16(h^{\rm R}_1)^2-
25{\bf h}\cdot {\bf h}]({\bf h}\cdot {\bf h})^{1/2}}
=-\frac{4a_1}{[9a_1^2+25(a_2^2+a_3^2)]^{1/2}(a_1^2+a_2^2+a_3^2)^{1/2}},
\label{A16}
\end{eqnarray}
we find that $(PVJ)^2=I$, and can thus identify $PVJ$ with $C$. With these specific values of $a$ and $b$, we note that $J$ is a positive, Hermitian operator. The reason why we needed to introduce the extra $J$ operator is that $V$ is not uniquely defined. Specifically, from (\ref{E8b}) we identified $V=B^{\dagger}B$. However, given any $J$ that commutes with $H$ we could have written (\ref{E8b}) in the form 
$B^{\dagger}BJHJ^{-1}B^{-1}(B^{\dagger})^{-1}=H^{\dagger}$, and then identified $V=B^{\dagger}BJ$. The overall normalization of $V$ is thus not fixed by (\ref{E8b}), and we can therefore use this freedom to construct a $C$ operator that not only commutes with $H$ but which is normalized to $C^2=I$ as well.

{}
\end{document}